\documentclass[11pt,a4paper]{article}
\usepackage{amstex,amsthm,amssymb,epsfig}

\def\At{{\widetilde A}}
\def\Bt{{\widetilde B}}
\def\Ct{{\widetilde C}}
\def\Dt{{\widetilde D}}
\def\C|{{\mathbb C} \,}
\def\B|{{\mathbb B} \,}
\def\S|{{\mathbb S} \,}
\def\G|{{\mathbb G} \,}
\def\N|{{\mathbb N} \,}
\def\F|{{\mathbb F} \,}
\def\K|{{\mathbb K} \,}
\def\BT{{\widetilde{\mathbb B}} \,}
\def\CT{{\widetilde{\mathbb C}} \,}
\def\Ft{{\widetilde F}}
\def\Rt{{\widetilde R}}
\def\Tt{{\widetilde T}}

\def\tens{\mathop\otimes\limits}

\def\sg{\sigma}
\def\sgm{\sigma^-}
\def\sgp{\sigma^+}
\def\sgz{\sigma^z}

\def\eps{\varepsilon}

\def\tr{\operatorname{tr}}

\newcommand{\bra}[1]{\langle\,#1\,|}
\newcommand{\ket}[1]{|\,#1\,\rangle}

\def\partialsur#1{\frac{\partial}{\partial#1}}
\def\Id{\mathrm{Id}}

\newtheorem{theorem}{Theorem}[section]
\newtheorem{prop}{Proposition}[section]

   \newtheorem{rem}{Remark}[section]
  \newtheorem{lemme}{Lemma}
\def\Proof{\medskip\noindent {\it Proof --- \ }}

\let\qed=\cqfd

\renewcommand{\theequation}{\thesection.\arabic{equation}}
\newcommand\beq{\begin{equation}}
\newcommand\enq{\end{equation}}

\def\beqa{\begin{eqnarray}}
\def\eeqa{\end{eqnarray}}
\def\ba{\begin{array}}
\def\ea{\end{array}}

\def\a{\alpha}
\def\b{\beta}

\def\det{\operatorname{det}}
\def\eps{\epsilon}
\def\la{\lambda}
\def\s{\sigma}
\def\sul{\sum\limits}
\def\pl{\prod\limits}
\def\lt({\left(}
\def\rt){\right)}
\def\pd #1{\frac{\partial}{\partial #1}}

\def\argum{\{\mu_j\},\{\la_k\}} 
\def\umarg{\{\la_k\},\{\mu_j\}}

\begin{document}

\begin{titlepage}
\begin{flushright}
LPENSL-TH-04/98\\
\end{flushright}
\par \vskip .1in \noindent

\begin{center}
{\LARGE Form factors of the XXZ Heisenberg spin-$1 \over 2$ finite chain}\\
\end{center}
  \par \vskip .3in \noindent

\begin{center}

      {\bf N. KITANINE$^{*}$,  J. M. MAILLET, V. TERRAS}
  \par \vskip .1in \noindent

{\sl  Laboratoire de Physique $^{**}$\\
Groupe de Physique Th\'eorique\\
       ENS Lyon, 46 all\'ee d'Italie 69364 Lyon CEDEX 07
       France}\\[0.6in]
\end{center}

\par \vskip .10in
\begin{center}
{\bf Abstract}\\
\end{center}

\begin{quote}
Form factors for local spin operators of the XXZ Heisenberg spin-$1 \over 2$ 
finite chain are computed. Representation theory of Drinfel'd twists for the quantum affine algebra ${\cal U}_q (\hat{sl}_2)$ in finite dimensional modules is used to calculate  scalar products of Bethe states (leading to Gaudin 
formula) and to solve the quantum inverse problem for local spin operators
in the finite chain. Hence, we obtain the representation of the {\mbox{n-spin}}
correlation functions in terms of expectation values
(in ferromagnetic reference state) of the operator 
entries of the quantum monodromy matrix satisfying Yang-Baxter algebra. 
This leads to the direct calculation 
of the  form factors of the XXZ Heisenberg spin-$1 \over 2$ finite chain as determinants 
of usual functions of the parameters of the model. A two-point correlation function for adjacent sites is also derived using similar techniques.
  
\end{quote}
\par \vskip 0.5in

\begin{flushleft}
\rule{5.1 in}{.007 in}\\
$^{*}$ {\small On leave of absence from the St Petersburg branch of the Steklov Mathematical Institute, Fontanka 27, St Petersburg 191011, Russia.}\\
$^{**}${\small URA 1325 du CNRS, associ\'ee \`a  l'Ecole
Normale Sup\'erieure de Lyon.}\\
{\small This work is supported by CNRS (France), the EC-TMR contract FMRX-CT96-0012}, and MAE fellowship 96/9804.\\
{\small email: nkitanin\symbol{'100}enslapp.ens-lyon.fr, maillet\symbol{'100}enslapp.ens-lyon.fr, vterras\symbol{'100}enslapp.ens-lyon.fr}\\[0.2 in]

July 1998
\end{flushleft}

\end{titlepage}

\section{Introduction}
\setcounter{equation}{0}
\label{sect:intro}

One of the most challenging problems in the theory of low dimensional quantum integrable models 
\cite{Bax1,F1,g,Smbk,bik,JM1}, 
after finding the spectrum and eigenstates of the  corresponding Hamiltonians, is to construct exact and manageable expressions of their form factors and correlation functions. This is a fundamental problem both to enlarge the range of applications of these models in the realm of condensed matter physics and to better understand their underlying mathematical structures. Until recently, only very few models were known for which correlation functions can be computed exactly. Typical examples are the Ising model (related to free fermions) \cite{Yang,WMTB,SMJ} and conformal field theories \cite{BPZ,KZ}.

\bigskip

Beyond these models, in the framework of integrable systems solvable by means of Bethe Ansatz \cite{l1,l2,FST,F1,Bax1,g,bik}, related to a Quantum Group structure \cite{Drin2,Jim1,Jim2,FRT} and associated to an R-matrix solving the Yang-Baxter equation, one can distinguish at present essentially two different but complementary approaches that have been designed to deal with this problem.

\bigskip

One of them relies on the study of  form factors and correlation functions of quantum integrable models directly in the infinite volume limit. The roots of this approach are twofold:

On the one hand, it comes from the study of analytic properties and bootstrap equations  for the factorized S-matrices and form factors of integrable quantum field theories in infinite volume \cite{Z2,Smbk}. Typical models here are the two-dimensional Sine-Gordon relativistic quantum field theory and the Non-Linear Schr\"odinger model. There it was realized that the set of equations satisfied by the form factors are closely related to the q-deformed Knizhnik-Zamolodchikov equations arising from representation theory of quantum affine algebras, and their q-deformed vertex operators \cite{FR,DJO,JKMQ,Sm2,Sm4,Sm3,TV,Var}.

On the other hand, it uses the Corner Transfer Matrix introduced by Baxter \cite{BaxCTM1,BaxCTM2,Bax1,Thacker} in the context of integrable models of statistical mechanics, and the relation of its spectra to characters of affine Lie algebras \cite{1Dsum}. Typical examples here are the six-vertex model and the XXZ Heisenberg spin-$1 \over 2$ infinite chain. In such models, using very plausible hypothesis about the representation of the Hamiltonian as a central element of the corresponding quantum affine algebra (here ${\cal U}_q ({\hat {sl_2}})$) in the infinite volume limit, the space of states is constructed in terms of highest weight  modules of ${\cal U}_q ({\hat {sl_2}})$ \cite{CORR,8v,JM1}, 
the combinatorial aspects of this construction being related to the theory of crystal bases \cite{DJO,(KMN)^2}. Form factors and correlation functions are then described in terms of q-deformed vertex operators, leading via bosonization \cite{KQS}, to integral formulas for them. As a result of its algebraic formulation, very parallel to the one in conformal field theory, there was a rapid development of this approach, leading to rather explicit expressions for correlators and their short distance behaviour \cite{JM1}.
 
It should be mentioned however, that the application of this method 
seems more difficult
for time or temperature dependent correlators, or, for quantum spin chains, if an external magnetic field is present, namely in situations where the infinite symmetry algebra is not as clearly identified as in the pure case. In these directions, one should cite \cite{Sm5,Sm6}, where in particular
a finite volume analysis of form factors has been undertaken. 

\bigskip

The other approach, described essentially in the book \cite{bik},  is based on the detailed analysis of the structure of Bethe eigenstates and in particular of their scalar products properties. One of the starting points of this approach is the Algebraic Bethe Ansatz (or Quantum Inverse Scattering) method \cite{FST,F1,Bax1,bik} and the derivation in this framework by Korepin of the Gaudin formula for the norm of Bethe eigenstates \cite{Kor}.  Then, to overcome the enormous combinatorial complexity due to the structure of Bethe eigenstates, the two  key ingredients of this method are on the one hand the so called ``dual fields approach'' \cite{k} and on the other hand the determinant expression for the partition function with ``domain wall'' boundary conditions \cite{Izer}. Using these auxiliary quantum ``dual fields'', determinant representations of correlation functions are obtained \cite{koriz,koriz1,efik,bik},  containing however vacuum expectation values of these auxiliary ``dual fields'', which cannot be eliminated in the final result. Hence, explicit expressions for the correlators cannot be obtained directly from this approach. Instead, the strategy is to embed these determinant formulas in systems of integrable integro-difference equations from which only large distance asymptotics of the correlation functions can be extracted from the resolution of (matrix)  Riemann-Hilbert problems.

Let us nevertheless note here, that in simpler models, in particular at so-called free-fermion points, such as the XX0 model or the Non-Linear Schr\"odinger model at infinite coupling constant, more explicit results can be obtained \cite{cikt1,cikt2}.

Although mainly restricted to the determination of these large distance asymptotics of the correlation functions, the very general formulation of this method allows one to apply it to a large variety of integrable models, and to correlation functions depending on time, temperature and eventually, in the case of spin chains, on an external magnetic field.

A more algebraic understanding of the Bethe Ansatz approach to correlation functions is certainly needed to avoid the combinatorial difficulties encountered in this method, in particular if one would like to obtain explicit expressions for the correlators, namely, without auxiliary ``dual fields''.

In this direction one should mention \cite{Skl1} where Gauss decomposition of operators was used within the Gaudin model to produce an explicit determinant formula for the norm of Bethe states, or also \cite{TV} where the Gaudin formula follows from semi-classical asymptotics of the q-deformed Knizhnik-Zamolodchikov equations.

\bigskip

The present state of the problem is such that, despite the great advances we just briefly described, algebraic derivation of form factors and correlation functions in an explicit and manageable setting, even for the most elementary models such as the XXZ Heisenberg spin-$1 \over 2$ {\em finite} chain, still poses a formidable problem.

\bigskip

The main motivation of this article is precisely to understand from a more algebraic point of view the Bethe Ansatz approach to correlation functions for finite systems, and to try eventually to relate the two above approaches in taking the thermodynamic limit. For that purpose, we will mainly concentrate on one of the a priori most elementary models in this context, the XXZ Heisenberg spin-$1 \over 2$ {\em finite} chain. We will show how to compute explicit determinant  formulas, namely in terms of usual functions of the parameters 
of the model and without any auxiliary ``dual fields'',  for the form factors  of local spin operators (i.e.\ their matrix elements between any two Bethe eigenstates) and for the adjacent sites two-point correlation function. In fact, we will also obtain this result for the completely inhomogeneous XXZ Heisenberg spin-$1 \over 2$ {\em finite} chain.

\bigskip

Our approach to form factors and correlation functions for this model decomposes into three main steps:
\begin{itemize}
\item[$i$.] we compute representations  for scalar products of an arbitrary Bethe eigenstate with any other state in terms of determinants of elementary functions of the parameters of the model.
\item[$ii$.] we solve the quantum inverse problem for the completely inhomogeneous XXZ Heisenberg spin-$1 \over 2$ {\em finite} chain, namely, we reconstruct the local spin operators at any site $i$ on the chain in terms of the elements of the quantum monodromy matrix of the chain.
\item[$iii$.] we combine these two results to obtain determinant formulas for the form factors of the local spin operators, and for an adjacent sites two-point 
correlation function.
\end{itemize}
 
The key ingredient of our method is the article \cite{maillet1}.  There a factorizing Drinfel'd twist $F$ was constructed and studied.
This twist is associated to the $N$-fold tensor product of spin-$1 \over 2$ (evaluation) representation of the  quantum affine algebra ${\cal U}_q ({\hat {sl_2}})$ associated to the completely inhomogeneous XXZ Heisenberg spin-$1 \over 2$ chain of length $N$. It has been shown in particular that the change of basis in quantum space of states generated by this twist $F$ is particularly convenient to study the structure of Bethe eigenstates in the framework of Algebraic Bethe Ansatz.  The main explanation of this is certainly the fact that the $F$-basis determines a completely symmetric presentation of the monodromy matrix operator for the (inhomogeneous) chain, such that the action of the symmetry group is trivial in this basis.

As a result, while creation and annihilation operators of Bethe states, namely the operator matrix elements of the quantum monodromy matrix, $B(\lambda)$ and $C(\lambda)$,  are in the original basis represented as huge sums, containing up to $2^N$ terms, each of them being a product of $N$ spin operators along the chain, their representations in this new $F$-basis simplify drastically. Indeed, in this basis, {\em they are given as sums of only $N$ terms, each of them being simply a local spin operator at some site $i$ of the chain, dressed by a pure tensor product of diagonal operators acting on the other sites} (see section 2 and \cite{maillet1}).

This means that the $F$-basis already solves the combinatorial problem of describing creation and annihilation operators of Bethe states. Moreover, we will show in this paper how it also solves the combinatorial problem  of describing Bethe eigenstates generated by products of creation operators $B(\lambda_k)$ on a reference (ferromagnetic) state. This will enable us to compute in an explicit way scalar products of a Bethe eigenstate with any other  state. The result is obtained for the completely inhomogeneous XXZ Heisenberg spin-$1 \over 2$ {\em finite} chain, as determinants of functions of the parameters of the model, and solves the above point ($i$).

Point ($ii$) is also solved by a careful study of the particularly simple structure of the quantum monodromy matrix in the $F$-basis. The reconstruction of local spin operators in terms of the operator matrix elements of the quantum monodromy matrix is then obtained in a basis-independent way.

Point ($iii$) uses only the combination of this two results and of some resummation formulas we will explain in the main text.

\bigskip

This article is organized as follows: in section 2 we recall the definition of  the inhomogeneous XXZ Heisenberg spin-$1 \over 2$ {\em finite} chain, and the algebraic ingredients we will use in the following, such as the quantum $R$-matrix and  the associated quantum monodromy matrix. Further, we describe  briefly formulas for the factorizing twist $F$ from \cite{maillet1} and its essential  properties to be  used in this article. In particular, we give there the expression of the quantum monodromy matrix in the $F$-basis. In section 3 we derive an explicit formula for the scalar product of an arbitrary Bethe state with any other state. Details of the proofs for this section are contained in the three appendices at the end of this article. Section 4 is devoted to the solution of the quantum inverse problem for the local spin operators. Finally, the main results of our work concerning form factors of the local spin operators  are presented in section 5. Conclusions and perspectives are given in section 6.

\section{The XXZ Heisenberg spin-$1 \over 2$ inhomogeneous finite chain}
\setcounter{equation}{0}
\label{sect:Heisenberg}

In this paper, we shall calculate form factors for the Heisenberg XXZ and XXX
spin-$1 \over 2$ chains of length $N$. The XXZ Heisenberg model is given by 
the following Hamiltonian:
\begin{equation}\label{HamXXZ}
  H_{\text{XXZ}}=J\sum_{m=1}^N \Big\{ \sigma^x_m \sigma^x_{m+1} +
  \sigma^y_m\sigma^y_{m+1} + \Delta(\sigma^z_m\sigma^z_{m+1}-1)\Big\},
\label{eq:H} 
\end{equation}
the particular case $\Delta=1$ corresponding to the XXX chain. We impose here
periodic boundary conditions.

\bigskip 

Our method is based on the Algebraic Bethe Ansatz \cite{FST,F1},
the central object of which 
is the quantum $R$-matrix. For the XXX and XXZ models it is of the
form 
\begin{equation}\label{mat-R}
   R(\lambda, \mu)=
      \begin{pmatrix}
         1 & 0 & 0 & 0\\
         0 & b(\lambda, \mu) & c(\lambda, \mu) & 0\\
         0 & c(\lambda, \mu) & b(\lambda, \mu) & 0\\
         0 & 0 & 0 & 1
      \end{pmatrix}
\end{equation}
where
\begin{align}\label{matbc}
      b(\lambda, \mu) & = \frac{\varphi(\lambda-\mu)}{\varphi(\lambda-\mu+\eta)},\\
      c(\lambda, \mu) & = \frac{\varphi(\eta)}{\varphi(\lambda-\mu+\eta)},
\end{align}
with the function $\varphi$ defined as
\begin{alignat}{3}\label{def-phi}
   \varphi(\lambda) & = \lambda         &  &\text{ in the XXX }&\text{case,}\\
   \varphi(\lambda) & = \sinh(\lambda)\ &  &\text{ in the XXZ }&\text{case.}    
\end{alignat}
The $R$-matrix is a linear operator in the tensor product of two two-dimensional
linear spaces $V_1 \otimes V_2$, where each $V_i$ is isomorphic to ${\bf C}^2$, and depends generically on two spectral parameters $\lambda_1$ and $\lambda_2$ associated to these two vector spaces. It is denoted by $R_{12} (\lambda_1, \lambda_2)$. Such an  $R$-matrix satisfies the Yang-Baxter equation,
\begin{equation}
  R_{12} (\lambda_1,\lambda_2)\ R_{13} (\lambda_1,\lambda_3)\ 
      R_{23} (\lambda_2,\lambda_3) = 
  R_{23} (\lambda_2,\lambda_3)\ R_{13} (\lambda_1,\lambda_3)\ 
      R_{12} (\lambda_1,\lambda_2),
\end{equation}
the unitary condition 
(provided $b(\lambda_1, \lambda_2) \ne \pm c(\lambda_1, \lambda_2)$),
\begin{equation}
  R_{12} (\lambda_1,\lambda_2)\ R_{21} (\lambda_2,\lambda_1) = {\mathbf 1},
\label{eq:eu}
\end{equation}
and the crossing symmetry relation,
\begin{equation}
  ({\gamma} \otimes {\mathbf 1})\ R_{12} (\lambda_1^s,\lambda_2)\ 
          ({\gamma} \otimes {\mathbf 1}) = 
    R_{21}^{t_1} (\lambda_2 ,\lambda_1 )\ \rho (\lambda_1,\lambda_2),
\label{eq:ecross}
\end{equation}
with $\rho (\lambda_1,\lambda_2)$ being a scalar function, and $\gamma$ a $2 \times 2$ matrix such that $\gamma^2 = {\mathbf 1}$, $\gamma^t  = \pm \gamma$, the upperscript $t_j$ meaning the usual transposition of matrices in the corresponding space $(j)$.
For the rational case,
\begin{equation*}
  \lambda_1^s = \lambda_1 - \eta,\quad 
  \gamma = \sigma^y,\quad 
  \rho (\lambda_1,\lambda_2) = 
       \frac{\lambda_1 - \lambda_2 - \eta}{\lambda_1 - \lambda_2},
\end{equation*}
and for the trigonometric case, 
\begin{equation*}
  \lambda_1^s = \lambda_1 - \eta + i \pi ,\quad 
  \gamma = \sigma^x ,\quad 
  \rho (\lambda_1,\lambda_2) = 
    \frac{\sinh(\lambda_1 - \lambda_2 - \eta)}{\sinh(\lambda_1 - \lambda_2)},
\end{equation*}
where $\sigma^x$ and $\sigma^y$ are the standard Pauli matrices.

Identifying one of the two linear spaces in the $R$-matrix 
with the two-dimensional Hilbert
space ${\cal H}_n$ of $SU(2)$ spin-$1 \over 2$ corresponding to the site $n$ of the chain, it is possible to construct the quantum $L$-operator of the model at  site $n$ as
\begin{equation}
 L_n(\lambda, \xi_n)=R_{0n}(\lambda, \xi_n),
\end{equation}
where $\xi_n$ is an arbitrary (inhomogeneity) parameter dependent on the site $n$. The subscripts mean here that $R_{0n}$ acts on the tensor product ${\bf{C}^2} \otimes {\cal H}_n$.
The quantum monodromy matrix of the total chain defined as the ordered product of $L$-operators is given by
\begin{equation}
T_0 (\lambda)\ \equiv\ T_{0, 1 \ldots N} (\lambda; \xi_1, \ldots ,\xi_N) =
      R_{0N}(\lambda, \xi_N)\ldots R_{01}(\lambda, \xi_1).
\end{equation}
It can be represented in the first space $0$ as a $2 \times 2$ matrix,
\begin{equation}\label{mat-monodromie}
   T (\lambda) = 
      \begin{pmatrix}
         A (\lambda) & B (\lambda)\\
         C (\lambda)& D (\lambda)
      \end{pmatrix}, 
\end{equation}
whose matrix elements 
$A (\lambda) \equiv A_{1 \ldots N} (\lambda; \xi_1, \ldots ,\xi_N)$, 
$B(\lambda) \equiv B_{1 \ldots N} (\lambda; \xi_1, \ldots ,\xi_N)$, 
$C(\lambda) \equiv C_{1 \ldots N} (\lambda; \xi_1, \ldots ,\xi_N)$, 
$D(\lambda) \equiv D_{1 \ldots N} (\lambda; \xi_1, \ldots ,\xi_N)$ 
are linear operators on the quantum space of states of the chain 
${\cal H}=\mathop\otimes\limits_{n=1}^{N}{\cal H}_n$. Their commutation relations are
given by the following relation on ${\bf C}^2 \otimes {\bf C}^2$:
\begin{equation} \label{commutation}
   R_{12}(\lambda, \mu)\ T_1(\lambda)\ T_2(\mu)  
      = T_2(\mu)\ T_1(\lambda)\ R_{12}(\lambda, \mu),
\end{equation}
with the usual tensor notations $T_1(\lambda)=T(\lambda) \otimes \Id$ and
$T_2(\mu)=\Id \otimes T(\mu)$.

The monodromy matrix satisfies moreover 
the following (crossing symmetry) relation leading to the 
definition of the quantum determinant:
\begin{equation}
  T_{0, 1 \ldots N} ({\lambda} ; \xi_1, \ldots, \xi_N)\ \gamma_0\ 
  T_{0, 1 \ldots N}^{t_0} ({\lambda}^s ; \xi_1, \ldots, \xi_N)\ \gamma_0 = 
  \rho ({\lambda}; \xi_1, \ldots, \xi_N)\ {\mathbf 1},
\label{eq:qdet}
\end{equation}
where $\rho ({\lambda}; \xi_1, \ldots, \xi_N) = 
\prod_{i = 1}^{N}\ \rho ({\lambda}, \xi_i)$, 
and $\gamma_0$ is the matrix $\gamma$ of eq.~\eqref{eq:ecross} acting in space $0$.

One also defines the transfer matrix $\mathcal{T} (\lambda)$ as the trace 
$A(\lambda)+D(\lambda)$ of the total monodromy matrix. 
Thanks to the Yang-Baxter equation and the invertibility of the $R$-matrix, the transfer matrices commute with each other for different values of the spectral parameter $\lambda$.  
For the homogeneous case where all parameters $\xi_i$ are equal, the Hamiltonian (\ref{eq:H}) can be obtained in terms of the transfer matrix by means of trace identities.

\bigskip

The Algebraic Bethe Ansatz, which deals with the problem of diagonalizing simultaneously  $\mathcal{T} (\lambda)$ for all values of $\lambda$, supposes the existence of a reference state
$\ket{0}$, called pseudo-vacuum, such that
\begin{equation}\begin{aligned}\label{eq-1-33}
   A(\lambda)\ket{0} & = a(\lambda)\ket{0},\\
   D(\lambda)\ket{0} & = d(\lambda)\ket{0},\\
   C(\lambda)\ket{0} & = 0,\\
   B(\lambda)\ket{0} & \ne 0.
\end{aligned}\end{equation}
For the XXX or XXZ model, the pseudo-vacuum is the completely ferromagnetic state with all the spins up, and  $a(\lambda)=1$,  
$d(\lambda)=\prod_{i=1}^N b(\lambda, \xi_i)$. Common eigenstates of the transfer matrices for different values of the spectral parameter $\lambda$ are obtained as successive actions of operators $B$ on the pseudo-vacuum 
$\prod_{j=1}^n B(\lambda_j)\ket{0}$, for any set of $n$ spectral parameters
$\{\lambda_j, 1 \le j \le n\}$ solution of Bethe equations
\begin{equation}\label{eq-bethe}
   r(\lambda_k)\prod\begin{Sb}j=1\\j \not= k\end{Sb}^n 
      \frac{b(\lambda_k, \lambda_j)}{b(\lambda_j, \lambda_k)} 
   = 1,                    \qquad 1 \le k \le n, 
\label{eq:bethe}
\end{equation}
with
\begin{equation}
r(\lambda) = \frac{a(\lambda)}{d(\lambda)}.
\label{eq:rl}
\end{equation}
The corresponding eigenvalue for the transfer matrix $\mathcal{T} (\mu)$ is then
\begin{equation}\label{theta}
   \tau(\mu, \{\lambda_j\})=
      a(\mu)\prod_{j=1}^n b^{-1}(\lambda_j, \mu) 
      + d(\mu)\prod_{j=1}^n b^{-1}(\mu, \lambda_j). 
\end{equation}

\bigskip

Let us now turn to the description of the key object we will use to compute scalar products of Bethe states and to solve the Quantum Inverse Problem for local spins, leading finally to the form factors formulas: the factorizing $F$-matrix associated to the above $R$-matrix.

\bigskip

The concept of factorizing $F$-matrices was defined 
in \cite{maillet1}, following the concept of 
twists introduced by Drinfel'd in the theory of Quantum Groups \cite{Drin2}. 
To be essentially self-contained we briefly recall here their 
main properties and refer to \cite{maillet1} for more details and proofs.

Due to the Yang-Baxter equation and to the unitarity of the $R$-matrix 
associated to the XXX and XXZ models, for any integer $n$ 
one can associate to any element $\sigma$ of the symmetric group $S_n$ 
of order $n$, a unique  $R$-matrix 
$R^{\sigma}_{1 \ldots n} (\xi_1, \ldots, \xi_n)$ 
constructed as some ordered product (depending on $\sigma$) of the elementary 
$R$-matrices $R_{ij} (\xi_i, \xi_j)$ (see \cite{maillet1}), such that
\begin{equation}
  R^{\sigma}_{1 \ldots n} (\xi_1, \ldots, \xi_n)\ 
  T_{0 , 1 \ldots n} (\lambda; \xi_1, \ldots, \xi_n) = 
  T_{0 , \sigma(1) \ldots \sigma(n)} 
       (\lambda; \xi_{\sigma(1)}, \ldots, \xi_{\sigma(n)})\ 
  R^{\sigma}_{1 \ldots n} (\xi_1, \ldots, \xi_n).
\end{equation}

A factorizing $F$-matrix associated to a given elementary $R$ matrix 
is an invertible matrix $F_{1 \ldots n} (\xi_1, \ldots, \xi_n)$ 
satisfying the following relation for any element $\sigma$ of $S_n$:
\begin{equation}\label{FfactoriseR}
  F_{\sigma(1) \ldots \sigma(n)} (\xi_{\sigma(1)}, \ldots, \xi_{\sigma(n)})\ 
  R^{\sigma}_{1 \ldots n} (\xi_1, \ldots, \xi_n) = 
  F_{1 \ldots n} (\xi_1, \ldots, \xi_n).
\end{equation}
In other words, such an $F$-matrix factories the corresponding $R$-matrix. 
Taking into account the fact that the parameters $\xi_i$ are in one to one 
correspondence with the vector spaces $V_i$, we can adopt simplified 
notations such that
\begin{align*}
  F_{1 \ldots n} (\xi_{1}, \ldots, \xi_n) &= F_{1 \ldots n},\\
  F_{\sigma(1) \ldots \sigma(n)} 
           (\xi_{\sigma(1)}, \ldots, \xi_{\sigma(n)}) &=     
      F_{\sigma(1) \ldots \sigma(n)},\\
  F_{1, 2 \ldots n} (\xi_{1} ; \xi_{2} \ldots,\ \xi_{n}) &= F_{1, 2 \ldots n}.
\end{align*}
An explicit formula for a triangular $F$-matrix corresponding to the XXZ model has been constructed in \cite{maillet1}. It reads for any integer $n$,
\begin{align}
  F_{1 \ldots n}  &= F_{2 \ldots n}\ F_{1, 2 \ldots n}, \\
              &= F_{n-1\, n}\ F_{n-2, n-1\, n} \ldots F_{1, 23 \ldots n},
\label{eq:deftf1}
\end{align}
where the partial $F$-matrices 
$F_{i, i+1 \ldots n}(\xi_{i}; \xi_{i+1}, \ldots ,\xi_n)$ 
are given in terms of the $R$-matrices as
\begin{equation}
  F_{i, i+1 \ldots n}(\xi_{i}; \xi_{i+1}, \ldots ,\xi_n) = e_i^{(11)} 
     + e_i^{(22)}\ R_{i, i+1 \ldots n}(\xi_{i}; \xi_{i+1}, \ldots ,\xi_n). \label{eq:frn}
\end{equation}
Here we have defined the partial $R$-matrices acting in 
$V_i \otimes \dots \otimes V_n$ as
\begin{equation}
  R_{i, i+1 \ldots n} (\xi_{i}; \xi_{i+1}, \ldots, \xi_{n}) = 
  R_{in} (\xi_i, \xi_n ) \ldots R_{i\, i+1} (\xi_i, \xi_{i+1}),   
\end{equation}
and $e_i^{(kl)}$ is the elementary matrix  $e^{(kl)}$ acting in space $i$, 
with matrix elements $e_{ab}^{(kl)} = \delta_{ak} \delta_{bl}$. 
The partial $F$-matrix $F_{0, 1 \ldots N}(\lambda; \xi_{1}, \ldots, \xi_N)$ 
has a useful expression as a $2 \times 2$ matrix in the first space $0$ 
in terms of elements of the quantum monodromy matrix:
\begin{eqnarray}
  F_{0, 1 \ldots N} (\lambda; \xi_1, \ldots, \xi_N)   
  = \left(
  \begin{array}{cc}
      {\mathbf 1} & {\mathbf 0} \\
      C_{1 \ldots N} (\lambda; \xi_1, \ldots, \xi_N)  & 
          D_{1 \ldots N} (\lambda; \xi_1, \ldots, \xi_N)
  \end{array}
  \right)_{[0]}.
\label{eq:22f}
\end{eqnarray}

Let us note here two important properties we will use in the following. The first one can be derived directly from the above relations between $R$, $T$ and $F$, leading to
\begin{multline*}
  F_{1 \ldots N} (\xi_1,  \ldots, \xi_N)\ 
  T_{0, 1 \ldots N}(\lambda; \xi_1,  \ldots, \xi_N)\ 
  F_{1 \ldots N}^{-1} (\xi_1,  \ldots, \xi_N)\ =\\
  =\ F_{\sigma(1) \ldots \sigma(N)} 
              (\xi_{\sigma(1)},  \ldots, \xi_{\sigma(N)})\ 
     T_{0, \sigma(1) \ldots \sigma(N)} 
              (\lambda; \xi_{\sigma(1)},  \ldots, \xi_{\sigma(N)}) \ 
     F_{\sigma(1) \ldots \sigma(N)}^{-1} 
              (\xi_{\sigma(1)}, \ldots, \xi_{\sigma(N)}).
\end{multline*}
Hence, it means that in the $F$-basis, the monodromy matrix $\Tt$ defined as
\begin{equation}
   {\Tt}_{0, 1 \ldots N} (\lambda; \xi_1,  \ldots , \xi_N)\ = \  
     F_{1 \ldots N} (\xi_1,  \ldots, \xi_N)
     T_{0, 1 \ldots N} (\lambda; \xi_1, \ldots , \xi_N)\ 
     F_{1 \ldots N}^{-1} (\xi_1,  \ldots, \xi_N),
\end{equation}
is totally symmetric under any simultaneous permutations of the lattice sites $i$ and of the corresponding inhomogeneity parameters $\xi_i$.

The second property, proved in \cite{maillet1}, is as follows: 
for the XXZ-$1 \over 2$ model, the  quantum monodromy operator is a 
$2 \times 2$ matrix with entries $A,\ B,\ C,\ D$ which are  obtained 
as sums of $2^{N-1}$ operators which themselves are products of $N$ local operators on the quantum chain. 
As an example, the $B$ operator is given as
\begin{equation}
   B_{1 \dots N} (\lambda) = 
    \sum_{i = 1}^{N}\ {\sigma}_i^-\ \Omega_i\ 
    + \sum_{i\ne j\ne k} {\sigma}_i^-\ ({\sigma}_j^-\ {\sigma}_k^+)\ 
               \Omega_{ijk}\ +\ higher\ terms,
\label{eq:B2N}
\end{equation}
where $\sgp$, $\sgm$ and $sgz$ are the standard Pauli matrices and 
the matrices $\Omega_i$, $\Omega_{ijk}$, are diagonal operators 
acting respectively on all sites but $i$, on all sites but $i, j, k$, 
and  the higher order terms involve more and more exchange spin terms like 
${\sigma}_j^-\ {\sigma}_k^+$. It means that the $B$ operator returns one spin 
somewhere on the chain, this operation being however dressed non-locally 
and with non-diagonal operators by multiple exchange terms of the type ${\sigma}_j^-\ {\sigma}_k^+$.

So, whereas these formulas in the original basis are quite involved and 
cannot be used in direct computations, their expressions in 
the $F$-basis simplify drastically. From \cite{maillet1} we have
\begin{align}
  {\Dt}_{1 \ldots N} (\lambda; \xi_1,  \ldots, \xi_N) &\equiv
   F_{1 \ldots N} (\xi_1,  \ldots, \xi_N)\ 
   D_{1 \ldots N} (\lambda; \xi_1,  \ldots, \xi_N)\ 
   F_{1 \ldots N}^{-1} (\xi_1,   \ldots, \xi_N) \nonumber\\ 
  & = \tens_{i = 1}^{N}\ 
        \left( 
        \begin{array}{cc}
           b (\lambda, \xi_i) & 0\\
           0 & 1
        \end{array}
        \right)_{[i]}.
\label{DbaseF}
\end{align}
The operator $\Bt$ representing the operator $B$ in the $F$-basis is given by
\begin{equation}\label{BbaseF}
    {\Bt}_{1 \ldots N} (\lambda) = \sum_{i = 1}^{N}\ {\sigma}_i^-\ 
                      c (\lambda, \xi_i)\ \tens_{j \ne i}\ 
                             \left( 
                             \begin{array}{cc}
                                     b (\lambda, \xi_j) & 0\\
                                     0 & b^{-1} (\xi_j, \xi_i)
                             \end{array}
                             \right)_{[j]}.
\end{equation}
Similarly we have for the operator ${\Ct}$,
\begin{equation}\label{CbaseF}
    {\Ct}_{1 \ldots N} (\lambda) = \sum_{i = 1}^{N}\ {\sigma}_i^+\ 
                      c (\lambda, \xi_i)\ \tens_{j \ne i}\ 
                             \left( 
                             \begin{array}{cc}
                                b (\lambda, \xi_j)\ b^{-1} (\xi_i, \xi_j) & 0\\
                                0 & 1
                             \end{array}
                             \right)_{[j]},
\end{equation}
and the operator $\At$ can be obtained from quantum determinant relations \eqref{eq:qdet}.

\bigskip
 
We wish first to stress that the operators ${\At},\ {\Bt},\ {\Ct},\ {\Dt}$ 
satisfy the same quadratic commutation relations as 
$A,\ B,\ C,\ D$. Second, each of the operators ${\Bt}$ and ${\Ct}$ is reduced 
to an elementary sum on the sites of the chain of the corresponding spin 
operator at each site {\em dressed diagonally}, which is to be compared to 
their expressions in the original basis where they are given as sums of 
$2^N$ terms involving much more complicated operators.

It really means that the factorizing $F$-matrices we have constructed solve 
the combinatorial problem induced by the non-trivial action of the permutation 
group $S_N$ given by the $R$-matrix. In the $F$-basis the action of the 
permutation group on the operators ${\At},\ {\Bt},\ {\Ct},\ {\Dt}$ is trivial.  
Moreover the operator ${\At}\ +\ {\Dt}$ which contains the Hamiltonian of the 
model together with the series of conserved quantities, is now a quasi-bi-local 
operator.

Further, it can be shown that the pseudo-vacuum state is left invariant, namely, it is an eigenvector of the total $F$-matrix with eigenvalue $1$. Hence, in particular, the Algebraic Bethe Ansatz can be carried out also in the $F$-basis. For the scalar products of the quantum states of the model we have
\begin{align}
  \bra{0}\ C (\lambda_1) \ldots C (\lambda_n)\ 
        &B (\lambda_{n+1}) \ldots B (\lambda_{2n})\ \ket{0} =\nonumber \\ 
   &= \bra{0}\ {\Ct} (\lambda_1) \ldots {\Ct} (\lambda_n)\ 
    {\Bt} (\lambda_{n+1}) \ldots {\Bt} (\lambda_{2n})\ \ket{0}.
\label{eq:sp}
\end{align}
Hence, thanks to these very simple expressions, a direct computation 
of Bethe eigenstates and of their scalar products in this $F$-basis 
is made possible, while it was completely hopeless in the original basis. 
There, only commutation relations between the operators $A,\ B,\ C,\ D$ 
can be used, leading (see \cite{bik}) to very intricate sums over partitions.

\bigskip

We now end this section with some useful formulas making the computation of the $F$-matrices simpler, namely the expressions of the partial $F$-matrices in the $F$-basis. This will help us to solve the quantum inverse problem for the local spin operators.

\bigskip

Factorizing $F$-matrices are given in (\ref{eq:deftf1}) as an ordered 
product of partial $F$-matrices like 
$F_{1, 2 \ldots n}$. 
These object are constructed in terms of the $R$-matrix 
$R_{1, 2 \dots n}$. 
However this quantity is highly non trivial to compute explicitly, 
since it involves in fact sums of $2^{n-1}$ terms. 
In contrast, the partial $F$-matrices in the $F$-basis can be obtained explicitly, while they also lead to the construction of factorizing 
$F$-matrices $F_{12 \ldots n}$. We have (using again simplified notations)
\begin{align}
    F_{1\ldots n}  &= {\Ft}_{1, 2 \ldots n} \ 
                   F_{2 \ldots n}\nonumber\\       
        &= {\Ft}_{1, 2 \ldots n} \ {\Ft}_{2, 3 \ldots n}  \ldots 
                        F_{n-1\, n},
\end{align}
and the partial $F$-matrix ${\Ft}_{1, 2 \ldots n}$ 
reads as a $2 \times 2$ matrix in the first space $1$:
\begin{align}
    {\Ft}_{1, 2 \ldots n}(\xi_1; \xi_2,  \ldots, \xi_n) &=
         F_{2 \ldots n}  (\xi_2,  \ldots, \xi_n)\ 
         F_{1, 2 \ldots n} (\xi_1; \xi_2, \ldots, \xi_n)\ 
         F_{2 \ldots n}^{-1}(\xi_2,  \ldots, \xi_n)\\
    &= \left(
       \begin{array}{cc}
          {\mathbf 1} & {\mathbf 0} \\
          {\Ct}_{2 \ldots n} (\xi_1; \xi_2,  \ldots, \xi_n) & 
               {\Dt}_{2 \ldots n}  (\xi_1; \xi_2,  \ldots, \xi_n)
       \end{array}
       \right)_{[1]}.
\label{Ft-mat}
\end{align}
It is a very simple object to compute from the formulas of this section. Hence using the $F$-basis we have also obtained a more explicit and elementary formula for the $F$-matrix itself.

\section{Scalar products of Bethe states and the Gaudin formula}
\setcounter{equation}{0}
\label{sect:pr-sc}

In this section we  calculate the following  scalar products of states constructed by the action
of the operators $B(\lambda)$ on the pseudo-vacuum, 
\begin{equation}\label{ps0}
  S_n(\{\mu_j\},\{\la_k\}) = 
  \bra{0}\ \prod_{j=1}^n C (\mu_j) \ 
        \prod_{k=1}^n B (\lambda_k)\ \ket{0} ,
\end{equation}
when one of the sets of parameters,
for example $\{\la_k\}$, 
is a solution of Bethe equations. Hence the state $\prod_{k=1}^n B(\la_k)\ket{0}$ is supposed to be an eigenvector
of the transfer matrix,
\begin{equation}
 \left(A(\mu)+D(\mu)\right)\pl_{k=1}^n 
 B(\la_k)\ket{0}=\tau(\mu,\{\la_k\})\pl_{k=1}^n B(\la_k)\ket{0},
\end{equation}
with the eigenvalue
\begin{equation}
 \tau(\mu,\{\la_k\})=a(\mu)\pl_{k=1}^n b^{-1}(\la_k,\mu)
           +d(\mu)\pl_{j=1}^n b^{-1}(\mu,\la_k).
 \label{tau}
\end{equation}
We will prove the following theorem:
\begin{theorem}
  Let  $\{\la_1,\dots,\la_n\}$ be a solution of Bethe equations
  \begin{equation}
    \frac{d(\la_j)}{a(\la_j)}\pl_{k\neq j}
    \frac{b(\la_k,\la_j)}{b(\la_j,\la_k)}=1,\qquad 1\le k \le n\nonumber,
  \end{equation}
  and $\{\mu_1,\dots,\mu_n\}$
  be an  arbitrary set of parameters. Then the scalar product \eqref{ps0} 
  can be represented as a ratio
  of two determinants
  \begin{equation}
    S_n(\{\mu_j\},\{\la_k\})=S_n(\{\la_k\},\{\mu_j\})
    =\frac{\mathrm{det} T(\{\mu_j\},\{\la_k\})}
          {\mathrm{det} V(\{\mu_j\},\{\la_k\})},
    \label{theorem1}
  \end{equation}
  of the following $n \times n$ matrices $T$ and $V$:  
  \begin{equation}
    T_{ab}=\pd{\la_a}\tau(\mu_b,\{\la_k\}),
    \qquad V_{ab}=\frac 1{\varphi(\mu_b-\la_a)},\qquad 1\le a,b \le n.
    \label{tv}
  \end{equation}
  \label{theor} 
\end{theorem}
\Proof
 Let us first note that the computation of the derivatives in the matrix $T$ 
and of the determinant of the matrix $V$ 
gives a formula obtained in \cite{s1}. However, the proofs 
proposed in \cite{s1,s2} are quite complicated and use some recursion
relations for the scalar product or the dual field representation. Here
we  give a direct proof of this formula for XXX and XXZ models.

The usual approach to the scalar product developed
in \cite{Kor,koriz,koriz1} is based  
on the commutation relations between the matrix elements of the monodromy matrix
(operators $A(\lambda)$, $B(\lambda)$, $C(\lambda)$ and 
$D(\lambda)$). It leads to the recursion relations for the scalar products.  
Instead of it we use the explicit representations for these operators in the $F$-basis.
Indeed, as the vacuum vector is invariant under the action of the operator $F$, 
the scalar product (\ref{ps0})    
can be rewritten in terms of the operators in the $F$-basis,
\begin{equation}\label{ps1}
 S_n = 
 \bra{0}\ \prod_{j=1}^n \Ct (\mu_j) \ 
       \prod_{k=1}^n \Bt (\lambda_k)\ \ket{0}.
\end{equation}
%

To perform the computation, it is convenient first to change the normalization
of the operators $B(\la)$ and $C(\la)$:
\begin{equation}
  \B| (\lambda)=\frac{B(\lambda)}{d(\lambda)},
  \qquad
  \C| (\lambda)=\frac{C(\lambda)}{d(\lambda)}.
\end{equation}
We thus want to calculate the ``renormalized'' scalar product in the $F$-basis
\begin{equation}
 \label{prodscF}
 \S|_n = \bra{0}\
 \CT(\mu_n) \ldots \CT(\mu_1)\
 \BT(\lambda_1) \ldots \BT(\lambda_n)\ \ket{0},
\end{equation}
in which we suppose $\{\la_k\}$ to be a solution of Bethe equations.

The idea is to insert in the scalar product  complete sets of states
$\ket{i_1,\dots,i_m}$ 
beyond each operator $\C| (\lambda)$ , where we denote by
$\ket{i_1,\dots,i_m}$ the state with $m$ spins down 
in the sites $i_1,\dots,i_m$ and
with $N-m$ spins up in the other sites. We are thus led to consider
the intermediate functions
\begin{equation}
 \G|^{(m)}(\{\la_k\},\mu_1,\dots,\mu_m,i_{m+1},\dots,i_n)=
    \bra{i_{m+1},\dots,i_n}\ \CT(\mu_m) \ldots \CT(\mu_1)\
    \BT(\lambda_1) \ldots \BT(\lambda_n)\ \ket{0},
 \label{interfun}
\end{equation}
the last one being the scalar product,
\[\G|^{(n)}(\{\la_k\},\mu_1,\dots,\mu_n)=\S|_n.\]

There is actually a very simple recursion relation between these functions,
\begin{multline}
  \G|^{(m)}(\{\la_k\},\mu_1,\dots,\mu_m,i_{m+1},\dots,i_n)=
    \sum\limits_{j\neq i_{m+1},\dots, i_n}
  \bra{i_{m+1},\dots,i_n}\CT(\mu_m)\ket{j,i_{m+1},\dots,i_n}\times\\
  \times\G|^{(m-1)}(\{\la_k\},\mu_1,\dots,\mu_{m-1},j,i_{m+1},\dots,i_n),
  \label{recur1}
\end{multline}
where the matrix elements of the operator $\CT(\mu_m)$ 
can be easily calculated in the $F$-basis:
\begin{equation}
 \bra{i_{m+1},\dots,i_n}\CT(\mu)\ket{j,i_{m+1},\dots,i_n}=\frac {\varphi(\eta)}
                {\varphi(\mu-\xi_j)}\pl_{a\neq j} b^{-1}(\xi_j,\xi_a)
 \pl_{l=m+1}^n\Bigr( b(\mu,\xi_{i_l})b(\xi_j,\xi_{i_l})\Bigl).\label{Celem}
\end{equation}
The function $\G|^{(0)}$, defined as
\[\G|^{(0)}(\{\la_k\},i_{1},\dots,i_n)=\bra{i_1,\dots,i_n}
\prod_{k=1}^n \BT (\lambda_k)\ \ket{0},\]
is closely related to the partition function of the six-vertex model 
with domain wall
boundary conditions, which was initially given in \cite{Izer}. 
In Appendix A we compute directly  this partition
function using the  
$F$-basis representation for the operators $B(\la)$ and $C(\la)$. The function $\G|^{(0)}$ is then calculated in Appendix B:
\begin{equation}\label{G0}
\G|^{(0)}(\{\la_\a\},i_{1},\dots,i_n)= 
     \frac{\pl_{\alpha=1}^n\pl_{k=1}^n \varphi(\lambda_\a-\xi_{i_k}+\eta)}
     {\pl_{j>k} \varphi(\xi_{i_k}-\xi_{i_j})
     \pl_{\alpha<\beta} \varphi(\lambda_{\b}-\lambda_{\a})
     }\det \mathcal{N}(\lbrace \lambda_{\alpha} \rbrace, \lbrace \xi_{i_j}\rbrace)\ ,
\end{equation}
where the $n \times n$ matrix $\mathcal{N}(\lbrace \lambda_{\alpha} \rbrace, \lbrace \xi_{i_j}\rbrace)$  is 
defined by 
\begin{equation}
 \label{matrXXX1}
   \mathcal{N}_{\alpha j} = \frac{\varphi(\eta)}
   {\varphi(\lambda_{\alpha}-\xi_{i_j}+\eta) 
    \varphi(\lambda_{\alpha}-\xi_{i_j})} ,\qquad 1\le\alpha , j\le n.
 \end{equation}
One should now perform the summation in the relation~\eqref{recur1} 
and compute successively the functions
$\G|^{(m)}$. It can be done using some identities for rational functions.  
Detailed calculations are given
in Appendix C. One finally obtains, both in the XXX and XXZ cases,
\begin{equation}
  \S|_n(\{\mu_j\},\{\la_\a\})=\G|^{(n)}(\{\la_\a\},\mu_1,\dots,\mu_n)=
  \frac {\det H(\{\la_\a\},\{\mu_j\})}
  {\pl_{j>k} 
  \varphi(\mu_k-\mu_j)
     \pl_{\alpha<\beta} \varphi(\lambda_{\b}-\lambda_{\a})},
\label{slavXXX}
\end{equation}
where the matrix elements of the $n \times n$  matrix $H(\{\la_\a\},\{\mu_j\})$ are 
\begin{equation}
  H_{a b}=\frac{\varphi(\eta)}{\varphi(\la_a-\mu_b)}
  \Bigl( r(\mu_b)\pl_{m\neq a}\varphi(\la_m-\mu_b+\eta)
     -\pl_{m\neq a}\varphi(\la_m-\mu_b-\eta) \Bigr).
\label{matrslav1}
\end{equation}
 This formula was originally obtained in \cite{s1}. It can be rewritten in a very simple form
in terms of eigenvalues of the transfer matrix \eqref{tau},
\begin{equation}
  S_n(\{\mu_j\},\{\la_\a\})=
    \frac{\pl_{\a =1}^n\pl_{j=1}^n\varphi(\mu_j-\la_\a)}{\pl_{j>k} 
  \varphi(\mu_k-\mu_j)
     \pl_{\alpha<\beta}\varphi(\lambda_{\b}-\lambda_{\a})}
     \det T(\{\mu_j\},\{\la_\a\}),
\label{withtau}
\end{equation}
where the matrix $T$  is a Jacobian,
\[T_{a b}=\pd{\la_a}\tau(\mu_b,\{\la_\a\}).\]
Using a well known formula, 
\begin{equation}
  \det V=  \frac{\pl_{a<b}\varphi(\la_a-\la_b)\pl_{j<k}\varphi(\mu_k-\mu_j)}
                {\pl_{k=1}^n\pl_{a=1}^n\varphi(\mu_k-\la_a)},
\end{equation}
with the matrix $V$  defined by (\ref{tv}),
one can express the coefficient in \eqref{withtau} as a determinant
and obtain finally the representation~\eqref{theorem1}. 

One should also mention that one can suppose from the beginning $\bra{0}\prod_{\a=1}^n C(\la_\alpha)$ to be
a Bethe state (instead of $\prod_{\a=1}^n B(\la_\a)\ket{0}$),
 make almost the same calculations, and obtain the following  result:
\begin{equation}
\S|_n(\{\la_\a\},\{\mu_j\})=\S|_n(\{\mu_j\},\{\la_\a\}).
\end{equation}  
\qed

\bigskip

It can be easily seen that taking the limit 
$\mu_a\rightarrow\la_a,\ a=1,\dots,n$,
in the expression \eqref{slavXXX}, 
 one obtains a very nice proof of the Gaudin formula for
the square of the norm of the Bethe wave function,
initially proved by Korepin \cite{Kor},
\begin{align}
  \N|_n &\equiv \bra{0}\ \prod_{j=1}^n \C| (\la_j) \ 
           \prod_{k=1}^n \B| (\la_k)\ \ket{0}\ ,\nonumber\\
        &= \varphi^n(\eta) \prod_{\alpha\ne\beta} \frac{\varphi(\la_\alpha- \la_\beta+\eta)}
{\varphi(\la_\alpha- \la_\beta)}
            \det \Phi'(\{\la_\a\}),
\label{gaudin}
\end{align}
where $\Phi'$ is a $n \times n$ matrix the elements of which are given by
\begin{align}
  \Phi_{ab}' &= -\partialsur\la_b 
                       \ln\biggl(r(\la_a)\prod_{k=1 \atop k\ne a}^n
                       \frac{b(\la_a, \la_k)}{b(\la_k, \la_a)}
                       \biggr).
\label{matrix-gaudin}
\end{align}

\section{Quantum inverse scattering problem for local spins and correlation
         functions}
\setcounter{equation}{0}

In the previous section we have calculated the scalar  products of Bethe states using the $F$-basis. Our purpose is now to 
compute form factors of local spin operators. One possibility is to write them in the $F$-basis,
which allows us to perform their  calculation in 
the same way as for the scalar products.
There is however a more simple and instructive way to proceed, which consists in
solving the quantum inverse scattering problem for the local spin operators,
that is expressing them only in terms of the operator entries $A$, $B$,
$C$, $D$ of the quantum monodromy matrix of the model.

\subsection{Local spin operators in the $F$-basis and quantum inverse problem}

As we just said, the  first possible way to compute the form factors 
is to  express the local spin operators in the $F$-basis.
Note that it can be directly done
for spin operators at the first or the last site  of the chain. For example,
due to the recursion relation 
$F_{1\ldots N}=\Ft_{1, 2\ldots N}\ F_{2\ldots N}$
and to the very simple form~\eqref{Ft-mat} of $\Ft_{1, 2\ldots N}$, a direct computation of a product of $2\times 2$ matrices in the
space $1$ gives the value for  $\sgm_1$ in the $F$-basis,
\begin{equation}
     F_{1\ldots N}\ \sgm_1\ F_{1\ldots N}^{-1} = 
             \Dt_{1\ldots N} (\xi_1; \xi_1,\ldots,\xi_N)\ \sgm_1.
\end{equation}
Similar expressions hold for $\sgp_N$, $\sgz_1$ and $\sgz_N$ in the $F$-basis.

Thus, taking into account that $\Dt$ is totally symmetric, for a given site $i$ of the chain this result can be simply translated
into the following formula:
\begin{equation}
     F_{i\ldots N 1\ldots i-1}\ \sgm_i\ 
         F_{i\ldots N 1\ldots i-1}^{-1} = 
                 \Dt_{1\ldots N} (\xi_i)\ \sgm_i ,
\end{equation}
where we used the short notation 
$\Dt_{1\ldots N}(\lambda) \equiv \Dt_{1\ldots N} (\lambda;\xi_1,\ldots,\xi_N)$ 
for $\lambda = \xi_i$.
Hence,  to calculate  the operator $ \sgm_i $ in the $F$-basis
 $F_{1\ldots N}\ \sgm_i\ F_{1\ldots N}^{-1}$,
one should evaluate the product of two ``permuted" $F$-matrices    
$F_{1\ldots N}\ F_{i\ldots N 1\ldots i-1}^{-1}$. 
It  can be considered  as   the expression in the $F$-basis of some
{\em propagator} $F_{i\ldots N 1\ldots i-1}^{-1}\ F_{1\ldots N}$, 
for which the following result holds:

\begin{lemme}\label{lemme:prop}
    Let $U_1^i$ be the propagator 
    $F_{i\ldots N 1\ldots i-1}^{-1}\ F_{1\ldots N}$
    from site $1$ to site $i$ of the chain. It can be written into the
    two following forms:
    \begin{align}
        U_1^i
        &= R_{i-1, i\ldots N 1\ldots i-2}\ldots 
           R_{2, 3\ldots N1}\ R_{1, 2\ldots N},
           \label{propagateur1}\\
        &= \prod_{\alpha=1}^{i-1} 
           \left( A_{1\ldots N} (\xi_{\alpha}) 
                + D_{1\ldots N} (\xi_{\alpha}) \right).  
           \label{propagateur2}
    \end{align}
\end{lemme}
    
\Proof
Equality \eqref{propagateur1} comes from the factorizing property for the
$F$-matrix~\eqref{FfactoriseR} for a cyclic permutation $\sg$~\cite{maillet1}:
\begin{equation}
    F_{\alpha\ldots N 1\ldots \alpha-1}\
    R_{\alpha-1, \alpha\ldots N 1\ldots \alpha-2}
    = F_{\alpha-1\ldots N 1\ldots \alpha-2}.
\end{equation}

Equality~\eqref{propagateur2} follows from the identity
\begin{equation} \label{ADR}
     A_{1\ldots N} (\xi_{\alpha}) + D_{1\ldots N} (\xi_{\alpha}) 
     = R_{\alpha, \alpha+1\ldots N 1\ldots \alpha-1} (\xi_{\alpha} ;
       \xi_{\alpha+1},\ldots, \xi_{\alpha-1}).
\end{equation}
The proof of~\eqref{ADR} is based on the remark that 
$R_{0 \alpha} (\xi_\alpha)$ is nothing but the permutation matrix 
$P_{0 \alpha}$ of the spaces $0$ and $\alpha$.
Writing $A_{1\ldots N} (\xi_{\alpha}) + D_{1\ldots N} (\xi_{\alpha})$
as a trace in the auxiliary space $V_0$, and making $P_{0 \alpha}$
act on every factor, we obtain, thanks to the cyclicity of the trace,
\begin{align*}
     A_{1\ldots N} (\xi_{\alpha}) + D_{1\ldots N} (\xi_{\alpha}) 
    &= \tr_0 \left(
              R_{0N}(\xi_{\alpha}) \ldots R_{0 \alpha+1}(\xi_{\alpha})
              \ P_{0\alpha}\
              R_{0 \alpha-1}(\xi_{\alpha}) \ldots R_{01}(\xi_{\alpha})
             \right), \\
    &= R_{\alpha \alpha-1}(\xi_{\alpha}) \ldots
       R_{\alpha 1}(\xi_{\alpha})\ R_{\alpha N}(\xi_{\alpha})
       \ldots R_{\alpha \alpha+1}(\xi_{\alpha}), \\
    &=  R_{\alpha,\alpha+1\ldots N 1\ldots \alpha-1} (\xi_{\alpha} ;
       \xi_{\alpha+1},\ldots, \xi_{\alpha-1}) ,
\end{align*}
which ends the proof of lemma~\ref{lemme:prop}.
Note that in~\eqref{propagateur1},~\eqref{propagateur2}, all factors commute 
with each others.
\qed

\begin{rem}\label{rm:inv-prop}
    The propagator $U_1^1$ through the whole chain being the identity,
    let us notice that
    \begin{equation}
        \prod_{\alpha=1}^{N} 
           \left( A_{1\ldots N} (\xi_{\alpha}) 
                + D_{1\ldots N} (\xi_{\alpha}) \right)
         = R_{N, 1\ldots N}\ldots R_{1, 2\ldots N}
         = 1.
    \end{equation}
    Thus, the inverse $\left( U_1^i \right)^{-1}$ of the propagator 
    on a part of the chain is nothing but 
    the propagator $U_{i+1}^1=\prod_{\alpha=i+1}^{N} 
           \left( A + D \right) (\xi_{\alpha})$ on the remaining part. 
\end{rem}

\begin{rem}
    The action of the propagator consists in shifting the beginning of the chain
    from site $1$ to site $i$. For an operator entry $X_{1\ldots N}$ of the
    monodromy matrix ($X=A$, $B$,
    $C$ or $D$), it means that
    \begin{equation}
        U_1^i \ X_{1\ldots N} = X_{i\ldots N1\ldots i-1} \ U_1^i,
    \end{equation}
    which, in terms of the monodromy matrix, can be written
    \begin{equation}
        U_1^i \ T_{0, 1\ldots N} = T_{0, i\ldots N1\ldots i-1} 
        \ U_1^i.
    \end{equation}
\end{rem}

The lemma~\ref{lemme:prop} allows us to obtain the value of any local spin
operator in the $F$-basis. For example, $\sgm_i$ for a given site $i$ of
the chain becomes
\begin{equation}\label{sgmF}
      F_{1\/\ldots\/ N} \ \sgm_i \ F_{1\/\ldots\/ N}^{-1}
      = \prod_{\alpha=1}^{i-1} 
           \bigl( \At + \Dt \bigr) (\xi_{\alpha}) \ \cdot\ 
        \Dt (\xi_i) \ \sgm_i \ \cdot\
        \prod_{\alpha=1}^{i-1} 
           \bigl( \At + \Dt \bigr)^{-1} (\xi_{\alpha}).
\end{equation}

Using this expression --- and similar ones for $\sgp_i$, $\sgz_i$ ---
it is possible to compute directly the corresponding form factor by the
same method  as for scalar products.
  
Moreover, as the expressions of the operators
$\Bt (\lambda)$ and $\Ct (\lambda)$ in the $F$-basis are quasilocal in
terms of the $\sgm_i$ or $\sgp_i$, it is a way to solve the quantum inverse problem for local spin
operators.
%
%
Indeed, a direct calculus in the $F$-basis
gives the identity
\begin{equation}
     \Dt (\xi_i) \ \sgm_i \ \bigl(\At+\Dt \bigr)(\xi_i) = \Bt (\xi_i) ,
\end{equation}
which, with~\eqref{sgmF}, leads to a new reconstruction of $\sgm_i$.
There exists a straightforward proof of this last formula, that we expose
in the next paragraph.

\subsection{Reconstruction of local spin operators}

In this paragraph, we prove an important result concerning the
reconstruction of any local spin operator in the inhomogeneous spin chain
in terms of elements of the monodromy matrix.

\begin{theorem}\label{thm:reconstr}
   Local spin operators at a given site $i$ of the inhomogeneous XXX or XXZ
   Heisenberg chain are given by
   \begin{align}
     \sgm_i &= \prod_{\alpha=1}^{i-1} \left( A + D \right) (\xi_{\alpha}) 
               \ \cdot\
               B(\xi_i) \ \cdot\
               \prod_{\alpha=i+1}^N \left( A + D \right) (\xi_{\alpha}),\\
     \sgp_i &= \prod_{\alpha=1}^{i-1} \left( A + D \right) (\xi_{\alpha}) 
               \ \cdot\
               C(\xi_i) \ \cdot\
               \prod_{\alpha=i+1}^N \left( A + D \right) (\xi_{\alpha}),\\
     \sgz_i &= \prod_{\alpha=1}^{i-1} \left( A + D \right) (\xi_{\alpha}) 
               \ \cdot\
               (A-D) (\xi_i) \ \cdot\
               \prod_{\alpha=i+1}^N \left( A + D \right) (\xi_{\alpha}).       
   \end{align}
\end{theorem}

The proof of this theorem is a straightforward consequence of the following 
lemma when $x_0$ is
respectively equal to $\sgm_0$, $\sgp_0$ and $\sgz_0$ :

\begin{lemme}\label{lemme:reconstr}
     Let $x_i$ be an operator acting on the quantum spin space $V_i$.
     We note $x_0$ the corresponding $2 \times 2$ matrix acting on the
     auxiliary space $V_0$. They are related by the identity
     \begin{equation}
        \tr_0 \left( x_0 \ R_{0, 1\ldots N} (\xi_i) \right)
        = \prod_{\alpha=1}^{i-1} \left( A + D \right)^{-1} (\xi_{\alpha}) 
           \ \cdot\
               x_i \ \cdot\
               \prod_{\alpha=1}^i \left( A + D \right) (\xi_{\alpha}) ,
     \end{equation}
     where the trace in the left hand side is taken on the matrix acting in $V_0$.
\end{lemme}

\Proof
Arguments used to prove the case $i=1$ are quite similar to those of 
lemma~\ref{lemme:prop}:
\begin{align*}
   \tr_0 (x_0\ R_{0, 1\ldots N} (\xi_1))
       &= \tr_0 (x_0\ R_{0N} (\xi_1)\ldots  R_{02} (\xi_1)\ P_{01}),\\
       &= x_1\ R_{1N} (\xi_1)  \ldots  R_{12} (\xi_1),\\
       &= x_1\ R_{1,2\ldots N} (\xi_1),
\end{align*}
and we conclude with lemma~\ref{lemme:prop}.

To prove the general case, let us notice that in the $F$-basis 
$\Rt_{0,1\ldots N} (\xi_i)$ is completely symmetric into the spaces
$1,\ldots, N$, which enables us to consider that the chain begins with site $i$:
\begin{align*}
   \tr_0 (x_0\ R_{0, 1\ldots N} (\xi_i))
       &= F_{1\ldots  N}^{-1} \    
          \tr_0 (x_0\ \Rt_{0, 1\ldots N} (\xi_i))\
          F_{1\ldots  N},\\
       &= F_{1\ldots  N}^{-1} \    
          \tr_0 (x_0\ \Rt_{0, i\ldots N 1\ldots i-1} (\xi_i))\
          F_{1\ldots  N},\\
       &= F_{1\ldots N}^{-1} \  F_{i\ldots N 1\ldots i-1}\  
          \tr_0 (x_0\ R_{0,i\ldots N 1\ldots i-1} (\xi_i))\
          F_{i\ldots N 1\ldots i-1}^{-1}\ F_{1\ldots  N},\\
       &= F_{1\ldots  N}^{-1} \  F_{i\ldots N 1\ldots i-1}\ 
          x_i\ \left( A + D \right) (\xi_i)\
          F_{i\ldots N 1\ldots i-1}^{-1}\ F_{1\ldots  N}.
\end{align*}
The value of the propagator 
$F_{i\ldots N 1\ldots i-1}^{-1}\ F_{1\ldots  N}$
is given by lemma~\ref{lemme:prop}, which concludes the proof of
lemma~\ref{lemme:reconstr}.
\qed

\subsection{General formula for correlation functions}

These results make it possible to write a general formula for any
$k$-point spin-spin correlation function between two Bethe states for the
inhomogeneous XXX-$1\over 2$ or XXZ-$1\over 2$ Heisenberg chain. 
Indeed, for any integer
$k$ and any subset $\{ i_j \}_{1\le j \le k}$ of $\{ 1,\ldots, N\}$,
with the convention $i_1< i_2 \ldots < i_k$, the correlation function
for spins at sites $i_1,\ldots, i_k$ between two Bethe states
$\bra{0}\ C(\mu_1) \ldots C(\mu_{n_1})$ and 
$B(\lambda_1)\ldots B(\lambda_{n_2})\ \ket{0}$
can be written into the following
form:
\begin{multline}
    \bra{0}\ C(\mu_1) \ldots C(\mu_{n_1})\ \sg^{\eps_1}_{i_1}\ 
               \sg^{\eps_2}_{i_2}\/
               \ldots\/\sg^{\eps_k}_{i_k}\
     B(\lambda_1)\ldots B(\lambda_{n_2})\ \ket{0}=\\
   = \prod_{\alpha=1}^{i_1-1} \prod_{j=1}^{n_1} b^{-1}(\mu_j, \xi_\alpha)\ \cdot
     \prod_{\alpha=i_k+1}^{N} \prod_{j=1}^{n_2} b^{-1}(\lambda_j, \xi_\alpha)\ 
       \times\\ 
   \times\ \bra{0}\ C(\mu_1) \ldots C(\mu_{n_1})\ \cdot\ 
          X^{\eps_1}_{i_1} (\xi_{i_1})\ \cdot
     \prod_{\alpha=i_1+1}^{i_2-1} \bigl( A+D \bigr) (\xi_\alpha)\ \cdot\
                   X^{\eps_2}_{i_2} (\xi_{i_2})\ \ldots\\
    \ldots  \prod_{\alpha=i_{k-1}+1}^{i_k-1} \bigl( A+D \bigr) (\xi_\alpha)\ 
                       \cdot\
                            X^{\eps_k}_{i_k} (\xi_{i_k})\ \cdot\
      B(\lambda_1)\ldots B(\lambda_{n_2})\ \ket{0},
\end{multline}
where $\eps_j,\ 1 \le j \le k$, takes the values $+,\ -$, or $z$,
$X^{\eps_j}$ being equal respectively to $C,\ B$ and $A-D$.

\bigskip

Hence we have reduced the problem of computing any correlation function of the XXZ model to a simpler problem written only in terms of the operator entries of the quantum monodromy matrix of the chain.

\bigskip

In the next section, we shall compute explicitly the form factors ($k=1$)
and the two-point correlation functions at adjacent sites ($k=2$ and $i_1=i_2-1$).

\section{Form factors}
\setcounter{equation}{0}

 We derive here explicit expressions for the form factors of 
the local spin operators
for the finite inhomogeneous XXX and XXZ chains. More precisely,
we calculate the matrix
elements of the operators $\s_m^+$,  $\s_m^-$ and  $\s_m^z$ between
two Bethe eigenstates. We also give an expression for the simplest
 correlation function of two spin operators at adjacent sites.
 
\subsection{Operators $\s_m^-$ and $\s_m^+$}

We begin with the calculation of the following one-point functions,
\begin{equation}
  F^-_n(m,\argum)=\bra{0}\ \pl_{j=1}^{n+1}C(\mu_j)\
                  \s_m^-\ \pl_{k=1}^n B(\la_k)\ \ket{0},
\end{equation}
and
\begin{equation}
  F^+_n(m,\umarg)=\bra{0}\ \pl_{k=1}^{n}C(\la_k)\
                  \s_m^+\ \pl_{j=1}^{n+1} B(\mu_j)\ \ket{0},
\end{equation} 
where $\{\la_k\}_n$ and $\{\mu_j\}_{n+1}$ are solutions of Bethe equations.
 Using the results of the previous sections we prove here that they admit
the following representations:
\begin{prop}
  For two Bethe states with spectral parameters
  $\{\la_k\}_n$ and $\{\mu_j\}_{n+1}$, the matrix element of the operator 
  $\s_m^-$ can be represented as a determinant,
  \begin{multline}
   F^-_n(m,\argum)=
    \frac{\phi_{m-1}(\{\mu_j\})}{\phi_{m-1}(\{\la_k\})}\,
    \frac{\pl_{j=1}^{n+1}\varphi(\mu_j-\xi_m+\eta)}
         {\pl_{k=1}^n\varphi(\la_k-\xi_m+\eta)}\times\\
   \times \frac{1}{\pl_{n+1\geq k>j\geq1} 
    \varphi(\mu_k-\mu_j)
     \pl_{1\leq\beta<\alpha\leq n} 
     \varphi(\lambda_{\b}-\lambda_{\a})}\det_{n+1} H^-(m,\argum),
  \label{ff-HN}
  \end{multline}
  where the coefficients $\phi_m(\{\la_k\})$ are
  \begin{equation}
    \phi_m(\{\la_k\})=\pl_{k=1}^n\pl_{j=1}^m b^{-1}(\la_k,\xi_j),
  \end{equation}   
  and the $(n+1)\times (n+1)$ matrix $H^-$ is defined as  
  \begin{align}
   &H^-_{a b}(m)=\frac{\varphi(\eta)}{\varphi(\mu_a-\la_b)}
    \biggl( a(\la_b)\pl_{j=1\atop{j\neq a}}^{n+1}\varphi(\mu_j-\la_b+\eta)-
         d(\la_b)\pl_{j=1\atop{j\neq a}}^{n+1}\varphi(\mu_j-\la_b-\eta)\biggr)
    \quad \text{for }\ b<n+1,\\
   &H^-_{a n+1}(m)=\frac{\varphi(\eta)}{\varphi(\mu_a-\xi_m+\eta)\varphi(\mu_a-\xi_m)}.
  \end{align}

  The matrix element $F^+_n(m,\umarg)$ of the operator $\s_m^+$ admits
  a similar representation,
  \begin{equation}
      F^+_n(m,\umarg)=\frac{\phi_m(\la_k)\,\phi_{m-1}(\la_k)}
               {\phi_{m-1}(\mu_j)\,\phi_{m}(\mu_j)}F^-_n(m,\argum).
  \end{equation}
  \end{prop}

Note that in the homogeneous limit $\xi_j=0,\ j=1,\dots,N$, 
the coefficients
$\phi_m(\{\la_k\})$ are expressed in terms of the total momentum $P$
of the state parameterized by $\{\la_k\}$, 
\[\phi_m(\{\la_k\})=\exp\{-iPm\},\]
with
$$
P = \frac{i}{N} \sul_{k=1}^n \ln (r(\lambda_k)).
$$

\Proof
The proof of these representations is rather straightforward.
As it was shown in previous section (theorem~\ref{thm:reconstr}) 
the local operator $\s_m^-$ can be expressed in terms 
of the transfer matrix and the operator $B(\xi_m)$ as
\[\s_m^-=\pl_{j=1}^{m-1} (A+D)(\xi_j)\ \cdot\ B(\xi_m)\ \cdot\ 
\pl_{j=m+1}^{N}(A+D)(\xi_j). \]
Since the Bethe states are eigenstates of the transfer matrix,
\begin{equation}
  \left( A(\xi_j)+D(\xi_j) \right) \pl_{k=1}^n B(\la_k)\ket{0}=
  \biggl( \pl_{a=1}^n b^{-1}(\la_a,\xi_j) \biggr) \pl_{k=1}^n B(\la_k)\ket{0},
\end{equation}
the product of the operators $A(\xi_j)+D(\xi_j)$ contributes
to the function $F^-_n(m,\argum)$ as a global factor:
\begin{equation}
  F^-_n(m,\argum)=\phi_m^{-1}(\{\la_k\})\,\phi_{m-1}(\{\mu_j\})\
  \bra{0}\ \pl_{j=1}^{n+1}C(\mu_j)\ B(\xi_m)\pl_{k=1}^n B(\la_k)\ \ket{0}.
\label{ffad}
\end{equation}
Here we used a simple property of the solutions of
 Bethe equations,
\[\pl_{k=1}^n\pl_{j=1}^N b^{-1}(\la_k,\xi_j)=1.\]
The right hand side of \eqref{ffad} thus reduces to a scalar product,
\begin{equation}
  F^-_n(m,\argum)=\phi_m^{-1}(\{\la_k\})\,
  \phi_{m-1}(\{\mu_j\})\
  S_{n+1}(\{\mu_j\},\{\xi_m,\la_1,\dots,\la_n\}),
\label{ff-spr}
\end{equation}
which, $\{\mu_j\}$ being a solution of
Bethe equations, can be computed by means of theorem~\ref{theor}. Hence
\begin{equation}
  F^-_n(m,\argum)=\phi_m^{-1}(\{\la_k\})\,
  \phi_{m-1}(\{\mu_j\})\ \frac{\det T_{n+1}(\{\mu_j\},
                            \{\xi_m,\la_1,\dots,\la_n\})}
  {\det V_{n+1}(\{\mu_j\},\{\xi_m,\la_1,\dots,\la_n\})},
\label{ff-vt}
\end{equation}
where $T$ and $V$ are $(n+1) \times (n+1)$ matrices defined similarly as 
in theorem~\ref{theor}.
Writing them explicitly one obtains the representation (\ref{ff-HN}).

 The form factor $F^+_n(m,\umarg)$ 
can be calculated analogously using the representation
for the operator $\s^+_m$ given by theorem~\ref{thm:reconstr}.
\qed 

\subsection{Operator $\s^z_m$}

We calculate here the matrix elements of the operator $\s^z_m$ between two Bethe states,
\[F^z_n(m,\argum)=\bra{0}\ \pl_{j=1}^{n}C(\mu_j)\ 
\s_m^z\ \pl_{k=1}^{n}B(\la_k)\ \ket{0}.\]
We prove the following representation for this one-point function:
\begin{prop}
  For two Bethe states with sets of spectral parameters
  $\{\la_k\}_n$ and $\{\mu_j\}_{n}$, the matrix element of the operator 
  $\s_m^z$ can be represented as a determinant,
  \begin{multline}
   F^z_n(m,\argum)=
    \frac{\phi_{m-1}(\{\mu_j\})}{\phi_{m-1}(\{\la_k\})}
    \pl_{j=1}^{n}\frac{\varphi(\mu_j-\xi_m+\eta)}
                      {\varphi(\la_j-\xi_m+\eta)}\times\\
   \times \frac{1}{{\pl_{j>k} 
     \varphi(\mu_k-\mu_j)
     \pl_{\alpha<\beta} \varphi(\lambda_{\b}-\lambda_{\a})}}
       \det_n \Bigl(H(\argum)-2P(m,\argum)\Bigr),
  \label{szXXX}
  \end{multline}
  where $H(\argum)$ is the same matrix as for 
  the determinant representation~\eqref{slavXXX} of the scalar product, 
  \[H_{a b}=\frac{\varphi(\eta)}
  {\varphi(\mu_a-\la_b)}\Bigl(a(\la_b)\pl_{j\neq a}\varphi(\mu_j-\la_b+\eta)
  -d(\la_b)\pl_{j\neq a}\varphi(\mu_j-\la_b-\eta)\Bigr)\qquad 1\le a,b \le n,\]
  and 
  $P(m,\argum)$ is a  matrix of rank one, 
  \begin{equation}
    P_{a b}(m)=\varphi(\eta) \frac{\pl_{k=1}^n\varphi(\la_k-\la_b+\eta)}
          {\varphi(\mu_a-\xi_m)\varphi(\mu_a-\xi_m+\eta)}
           \qquad 1\le a,b \le n.
  \end{equation}
\end{prop}

\Proof
It is a bit more complicated than for $\s^-$ or $\s^+$.
We use here a representation for $\s_m^z$ which follows directly 
from remark~\ref{rm:inv-prop} and theorem~\ref{thm:reconstr},
\begin{equation}
  \s_m^z=2\ \pl_{j=1}^{m-1}(A+D)(\xi_j)\ \cdot\ A(\xi_m)\ \cdot\
      \pl_{j=m+1}^{N}(A+D)(\xi_j) -I,
\end{equation}
which leads to
\[F^z_n(m,\argum)=
2\ \phi_m^{-1}(\{\la_k\})\,
\phi_{m-1}(\{\mu_j\})\ P_1(\xi_m,\argum)-S(\argum),\]
with
\begin{equation}
  P_1(\xi_m,\argum)=\bra{0}\ \pl_{j=1}^{n}C(\mu_j)\
  A(\xi_m)\ \pl_{k=1}^{n}B(\la_k)\ \ket{0}.
\end{equation}
Let us now compute the function $P_1$.
It can be done using a well-known formula for the action of the operator $A(\xi)$ on an arbitrary state:
\begin{align}
   A(\xi_m)\pl_{k=1}^{n}B(\la_k)\ket{0}&=
   \pl_{k=1}^{n}b^{-1}(\la_k,\xi_m)\pl_{k=1}^{n}B(\la_k)\ket{0}-
   \nonumber\\
  -\sul_{a=1}^N &\frac{\varphi(\eta)}{\varphi(\la_a-\xi_m)}
   \biggl(\pl_{k=1\atop{k\neq a}}^n   
   \frac{\varphi(\la_k-\la_a+\eta)}{\varphi(\la_k-\la_a)}
   \biggr)\,B(\xi_m)\pl_{k=1\atop{k\neq a}}^n B(\la_k)
   \ket{0}.
\end{align}
Hence $P_1$ reduces to a sum of scalar products, therefore to a sum of
determinants according to 
theorem~\ref{theor}. It can be rewritten as a single
determinant by means of the following formula for the determinant of the sum of two matrices one of which being of rank one. 
Indeed, if $\mathcal{A}$ is an arbitrary $n\times n$ matrix and  
$\mathcal{B}$ a rank one $n\times n$ matrix, 
the determinant of the sum $\mathcal{A}+\mathcal{B}$ is: 
\[\det(\mathcal{A}+\mathcal{B})=\det\mathcal{A}+\sul_{j=1}^n\det\mathcal{A}^{(j)},\]
where 
\begin{align*}
  \mathcal{A}^{(j)}_{a b}&=\mathcal{A}_{a b}\quad \text{for\ }\ b\neq j, \\
  \mathcal{A}^{(j)}_{a j}&=\mathcal{B}_{a j}.
\end{align*}
 
   Using this formula and also  the orthogonality of two  
 different Bethe states, one obtains the determinant representation (\ref{szXXX}).
\qed

\bigskip

Let us remark that in the homogeneous limit $\xi_j=0,\ j=1,\dots,N$, 
and for two identical Bethe eigenstates, 
the evident mean value of $\s^z_m$ can be easily derived from these 
representations. Indeed~\eqref{szXXX} and~\eqref{gaudin} yield
\begin{equation}
  s^z(n)\equiv
  \frac{\bra{0}\ \pl_{j=1}^{n}C(\la_j)\ \s_m^z\ \pl_{k=1}^{n}B(\la_k)\ \ket{0}}
       {\bra{0}\ \pl_{j=1}^{n}C(\la_j)\ \pl_{k=1}^{n}B(\la_k)\ \ket{0}}
  =\frac{\det(\Phi'(\{\la_j\})-2P'(\{\la_j\})}{\det\Phi'(\{\la_j\}))},
\end{equation}
where $\Phi'(\{\la_j\})$ is the Gaudin matrix given by \eqref{matrix-gaudin} and $P'(\{\la_j\})$
a matrix of rank one:
\[P'_{a b}=-\frac 1N \pd{\la_a}\ln r(\la_a).\]
Due to the following property of the Gaudin matrix,
\[\sul_{b=1}^n\Phi'_{a b}=-\pd{\la_a}\ln r(\la_a),\]
it can be rewritten as
\[s^z(n)=\det_n(I-2U),\qquad U_{a b}=\frac 1N,\]
which leads to the evident value of $s^z(n)$ ($U$ being a rank one matrix):
\[s^z(n)=1-\frac {2n}N.\]

\subsection{The correlation function of spins at adjacent sites}

The next quantity to compute is the simplest two-point function, 
the correlator of two spins 
at adjacent sites,
\begin{equation}
  F^{-+}_n(m,m+1,\argum)\equiv
  \bra{0}\ \pl_{j=1}^{n}C(\mu_j)\ \s_m^-\,
     \s_{m+1}^+\ \pl_{k=1}^{n}B(\la_k)\ \ket{0}.  
\end{equation}
As usually $\{\la_k\}$ and $\{\mu_j\}$ are supposed to be solutions of Bethe equations.
From theorem~\ref{thm:reconstr} this function can be written  only 
in terms of the operators $A$, $B$,
$C$ and $D$. As previously, Bethe states being eigenstates for the propagator and the propagator from the 
first to the last site being equal to identity,
we obtain a simple expression for this correlation function,
\begin{multline}
  F^{-+}_n(m,m+1,\argum)=
    \phi_{m-1}(\{\mu_j\})\,\phi^{-1}_{m+1}(\{\la_k\})\\
    \bra{0}\ \pl_{j=1}^{n}C(\mu_j)\ B(\xi_m)
    \,C(\xi_{m+1})\ \pl_{k=1}^{n}B(\la_k)\ \ket{0}.  
\end{multline}
 It can then be reduced to a sum of scalar products by means of 
the commutation relations
between the matrix elements of the monodromy matrix, namely,
%
\begin{equation}
   C(\xi_{m+1})\pl_{k=1}^n B(\la_k)\ket{0}=\sul_{a=1}^n 
   M_a\pl_{k=1\atop{k\neq a}}^n B(\la_k)\ket{0}
   +\sul_{a\neq b}M_{a b}B(\xi_{m+1})
    \pl_{k=1\atop{k\neq a,b}}^n B(\la_k)\ket{0},
\end{equation}
with the coefficients $M_a$ and $M_{a b}$ given by 
\begin{equation}
\begin{aligned}
  M_a &= \frac{\varphi(\eta)}{\varphi(\la_a-\xi_{m+1})} d(\la_a)
  \pl_{k\neq a}^n b^{-1}(\la_k,\xi_{m+1}) b^{-1}(\la_a,\la_k),\\
  M_{a b}&=-\frac{\varphi^2(\eta)}
  {\varphi(\la_a-\xi_{m+1})\varphi(\la_b-\xi_{m+1})}
  d(\la_a)b^{-1}(\la_a,\la_b)\pl_{k\neq a,b}^n 
  b^{-1}(\la_k,\la_b) b^{-1}(\la_a,\la_k).
\end{aligned}
\end{equation}
Hence
\begin{multline}
  F^{-+}_n(m,m+1,\argum)=
                \phi_{m-1}(\{\mu_j\})\,\phi^{-1}_{m+1}(\{\la_k\})\times\\
   \times \biggl(\sul_{a=1}^n M_a 
      S_n(\{\mu_j\},\{\xi_m,\la_1,\dots,\widehat{\la}_a,\dots,\la_n\})+\\ 
   +\sul_{a\neq b}M_{a b}
                            S_n(\{\mu_j\},\{\xi_m,\xi_{m+1},\la_1,\dots,
       \widehat{\la}_a,\widehat{\la}_b,\dots,\la_n\}) \biggr),
\end{multline}
where  
the hat means that the corresponding parameter is not present in the set. 
Since $\{\mu_j\}$ is a solution of Bethe equations, the scalar products
$S_n(\argum)$
can be represented as determinants of size $n$ according to 
theorem~\ref{theor}.
Therefore, two-point functions at adjacent sites are obtained as a sum
of determinants of size $n$.

\section{Conclusion}
\setcounter{equation}{0}
\label{sect:Conclusion}

In this article we have computed explicit determinant representations for form factors of the XXZ Heisenberg spin-$1 \over 2$ inhomogeneous finite chain. These determinants are simply given in terms of usual functions of the parameters of the model. Moreover, an adjacent sites correlator has also been determined using similar techniques. The knowledge of the factorizing $F$-matrices from \cite{maillet1} was an essential ingredient to achieve this goal.  It also sheds some new light on the algebraic structure underlying the Bethe Ansatz approach to correlation functions. In particular, multi-point correlators have been expressed in terms of expectation values (on the ferromagnetic reference state) of operator entries of the quantum monodromy matrix: this was the result of explicitly solving the quantum inverse problem for local spin operators at any site of the chain.

Let us stress also that this method allowed us to give a very direct proof of the scalar product formula between a Bethe eigenstate and an arbitrary state generated by the successive actions of the operators $B$. This formula is beautiful but very mysterious, since it involves the Jacobian of the eigenvalues of the transfer matrix with respect to the parameters of the Bethe states. Although the proof is very transparent, we do not know a satisfactory a priori explanation of it, and one feels that there should be a more direct understanding of this formula.

What remains to be done, at least for the form factors we have computed, is to describe their thermodynamic limit and to compare the obtained results to 
those 
following the approach \cite{JM1}. This will be done in a forthcoming publication, for the spontaneous magnetization of the XXZ chain.

Another interesting question concerns the higher spin Heisenberg models. To deal with these more general cases, one would like to have at disposal the analogue of the factorizing $F$-matrices, but here for the higher (fused) $R$-matrices.  This question is now under study.

\par \vskip .5in \noindent
{\bf \Large Acknowledgements.} 
We would like to thank A. Izergin for useful discussions and for his interest in this work.

\section*{Appendix A}
\renewcommand{\theequation}{A.\arabic{equation}}
\setcounter{equation}{0}


We give in this appendix the determinant representation for the  
partition function of the  six-vertex model with 
domain wall boundary conditions, initially obtained in \cite{Izer}, 
and propose a direct proof for it, using the $F$-basis.  

The partition function of the six-vertex model with domain wall boundary
conditions corresponds to a special case of scalar product for the
XXX or XXZ 1/2 spin chain:
\begin{multline}\label{part}
  Z_N \left( \lbrace \lambda_{\alpha} \rbrace, \lbrace \xi_j \rbrace \right) =
    \bigg\lbrace \prod_{j=1}^N \downarrow_j \bigg\rbrace
    \bigg\lbrace B_{1 \ldots N} (\lambda_1;\xi_1,\ldots,\xi_N)\ 
    B_{1 \ldots N}(\lambda_2;\xi_1,\ldots,\xi_N) \ldots \\
    \ldots
    B_{1 \ldots N}(\lambda_N;\xi_1,\ldots,\xi_N) \bigg\rbrace
    \bigg\lbrace \prod_{j=1}^N \uparrow_j \bigg\rbrace.
\end{multline}
It should be mentioned that the same partition function can be represented 
similarly
as a matrix element of the products of operators $C(\la)$:
\begin{multline}\label{Cpart}
  Z_N \left( \lbrace \lambda_{\alpha} \rbrace, \lbrace \xi_j \rbrace \right) =
    \bigg\lbrace \prod_{j=1}^N \uparrow_j \bigg\rbrace
    \bigg\lbrace C_{1 \ldots N} (\xi_1;\lambda_1,\ldots,\lambda_N)\ 
    C_{1 \ldots N}(\xi_2;\lambda_1,\ldots,\lambda_N) \ldots \\
    \ldots
    C_{1 \ldots N}(\xi_N;\lambda_1,\ldots,\lambda_N) \bigg\rbrace
    \bigg\lbrace \prod_{j=1}^N \downarrow_j \bigg\rbrace.
\end{multline} 
%
Recursion relations for this function were obtained in \cite{Kor}. It was shown that they define completely the function. 
The solution to these relations was given in~\cite{Izer} as a determinant  
both in the XXX and XXZ
cases. For the XXX case one has the following result: 
\begin{equation}
\label{partition}
   Z_N \left(\lbrace \lambda_{\alpha} \rbrace, \lbrace \xi_j\rbrace \right)
   = \frac{\prod_{j=1}^N\prod_{\alpha=1}^N 
              \left(\lambda_{\alpha}-\xi_j\right)}
          {\prod_{j>k} \left(\xi_k-\xi_j\right)
     \prod_{\alpha>\beta} \left(\lambda_{\alpha}-\lambda_{\beta}\right)
     }\det \mathcal{N}(\lbrace \lambda_{\alpha} \rbrace, \lbrace \xi_j\rbrace)\ ,
\end{equation}
where $\mathcal{N}(\lbrace \lambda_{\alpha} \rbrace, \lbrace \xi_j\rbrace)$ is the $N \times N$ matrix given by
\begin{equation}
 \label{matrXXX}
   \mathcal{N}_{\alpha j} = \frac{\eta}{(\lambda_{\alpha}-\xi_j+\eta) 
                                        (\lambda_{\alpha}-\xi_j)}.
\end{equation}
In the XXZ case, the expression is similar:
\begin{equation}
\label{partitionZ}
   Z_N \left(\lbrace \lambda_{\alpha} \rbrace, \lbrace \xi_j\rbrace \right)
   = \frac{\prod_{j=1}^N\prod_{\alpha=1}^N \sinh(\lambda_{\alpha}-\xi_j)}
          {\prod_{j>k} \sinh(\xi_k-\xi_j)
     \prod_{\alpha>\beta} \sinh(\lambda_{\alpha}-\lambda_{\beta})
     }\det \mathcal{N}^{\mbox{\tiny XXZ}}(\lbrace \lambda_{\alpha} \rbrace, \lbrace \xi_j\rbrace)\ ,
\end{equation}
 where $\mathcal{N}^{\mbox{\tiny XXZ}}(\lbrace \lambda_{\alpha} \rbrace, \lbrace \xi_j\rbrace)$ is the $N \times N$ matrix given by
\begin{equation}
\label{matrXXZ}
   \mathcal{N}^{\mbox{\tiny XXZ}}_{\alpha j} = 
     \frac{\sinh\eta}
          {\sinh(\lambda_{\alpha}-\xi_j+\eta)\sinh(\lambda_{\alpha}-\xi_j)}.
\end{equation}

We give now an explicit derivation of these representations, based on direct 
calculations in the $F$-basis. Indeed, using the expression of operator
$\Bt$ (or $\Ct$), one obtains a new recursion formula for the partition 
function, which corresponds to a development of the determinant 
in~\eqref{partition} or~\eqref{partitionZ}. Explicit calculations are 
performed here in the XXX case, but they are quite similar in the XXZ case.

More precisely, as the state 
$\big\lbrace \prod_{j=1}^N \uparrow_j \big\rbrace$
(respectively $\big\lbrace \prod_{j=1}^N \downarrow_j \big\rbrace$) is invariant
under the left-action of $F_{1\ldots N}$  
(respectively the right-action of
$F_{1\ldots N}^{-1}$), the formula~\eqref{part} can be directly
written in the $F$-basis:
$$Z_N = \bigg\lbrace\prod_{j=1}^N \downarrow_j\bigg\rbrace
\bigg\lbrace \Bt_{1 \ldots N}(\lambda_1;\xi_1,\ldots,\xi_N)
\ldots
\Bt_{1 \ldots N}(\lambda_N;\xi_1,\ldots,\xi_N) \bigg\rbrace
\bigg\lbrace \prod_{j=1}^N \uparrow_j \bigg\rbrace\ ,$$
and, using the expression~\eqref{BbaseF}, we make 
$\Bt_{1\ldots N}(\lambda_N;\xi_1,\ldots,\xi_N)$
act on the state $\left\lbrace \prod_{j=1}^N \uparrow_j \right\rbrace$,
in order to obtain a recursion relation for $Z_N$:
\begin{equation*}
   Z_N = \sum_{i=1}^N c(\lambda_N, \xi_i)
     \biggl( \prod_{j=1\atop j\not=i}^N b(\lambda_N, \xi_j) \biggr)
     \bigg\lbrace \prod_{j=1}^N \downarrow_j \bigg\rbrace
     \bigg\lbrace \Bt_{1 \ldots N}(\lambda_1) \ldots
     \Bt_{1 \ldots N}(\lambda_{N-1}) \bigg\rbrace
     \bigg\lbrace \bigg(\prod_{j=1\atop j\not=i}^N \uparrow_j \bigg)
     \left( \downarrow_i \right) \bigg\rbrace.
\end{equation*}
We thus extract, for each term $i$ of the sum, the action 
(which is now diagonal for  $(\sigma_i^-)^2 = 0$) on space $i$ of the 
other operators $\Bt_{1 \ldots N}(\lambda_{\alpha})$, $1 \le \alpha \le N-1$, which leads to an extra numerical factor:
\begin{multline*}
Z_N = \sum_{i=1}^N c(\lambda_N, \xi_i)
\biggl( \prod_{j\not=i} b(\lambda_N, \xi_j)\, b^{-1}(\xi_i, \xi_j) 
  \biggr) \times\\
\times\ \bigg\lbrace \prod_{j\not=i} \downarrow_j \bigg\rbrace
\bigg\lbrace \Bt_{1 \ldots i-1\, i+1 \ldots N}
(\lambda_1; \xi_1, \ldots ,\xi_{i-1}, \xi_{i+1}, \ldots ,\xi_N) \ldots\\
\ldots \Bt_{1 \ldots i-1\, i+1 \ldots N}
(\lambda_{N-1}; \xi_1, \ldots ,\xi_{i-1}, \xi_{i+1}, \ldots ,\xi_N) \bigg\rbrace
\bigg\lbrace \prod_{j\not=i} \uparrow_j \bigg\rbrace.
\end{multline*}
Indeed, as the number of operators $\Bt$ was formerly equal to the
number of sites on the chain, and as $(\sigma_j^-)^2 = 0$, the product
$\prod_{j=1}^N \sigma_j^-$ appears in all the non-zero
terms in the development of the product of $\tilde{B}$. 

We have obtained the following recursion formula for $Z_N$:
\begin{equation}
  Z_N \left(\lbrace \lambda_{\alpha} \rbrace_{1 \le \alpha \le N},
   \lbrace \xi_j \rbrace_{1 \le j \le N} \right) =
   \sum_{i=1}^N c(\lambda_N, \xi_i)
   \biggl( \prod_{j=1\atop j\not=i}^N b(\lambda_N, \xi_j)\,
   b^{-1}( \xi_i, \xi_j) \biggr)
   Z_{N-1}\left(\lbrace \lambda_{\alpha} \rbrace_{\alpha\not=N},
   \lbrace \xi_j \rbrace_{j\not=i} \right). \label{recZ}
\end{equation}
This corresponds actually to the last line development of the determinant
in the formula
\begin{equation*}
  Z_N \left(\lbrace \lambda_{\alpha} \rbrace, \lbrace \xi_j \rbrace\right) =
   \frac{\prod_{j=1}^N \prod_{\alpha=1}^N (\lambda_{\alpha} - \xi_j)}
   {\prod_{j>k}(\xi_k - \xi_j)
   \prod_{\alpha>\beta}(\lambda_{\alpha} - \lambda_{\beta})}
   \det\hat{\mathcal{N}}\ ,
\end{equation*}
where $\hat{\mathcal{N}}$ is the matrix obtained from $\mathcal{N}$ 
by adding to the last line $L_N$ the linear combination of the other lines
$\sum_{\beta=1}^{N-1} f_{\beta}L_{\beta}$, with coefficients
\begin{equation*}
  f_{\beta} = - \prod_{k=1}^N \frac{\lambda_{\beta} - \xi_k + \eta }
    {\lambda_N - \xi_k + \eta} \cdot
    \prod_{\alpha=1\atop \alpha\not=\beta}^{N-1}
    \frac{\lambda_N - \lambda_{\alpha}}{\lambda_{\beta} - \lambda_{\alpha}}.
\end{equation*}

 In fact, this development leads to the following recursion formula
for $Z_N$:
\begin{multline*}
  Z_N = \sum_{i=1}^N \frac{\eta}{\lambda_N - \xi_i + \eta}
    \Biggl(\prod_{k=1\atop k\not=i}^N
	\frac{\lambda_N - \xi_k}{\lambda_N - \xi_k + \eta}\Biggr)
    \Biggl(\prod_{k=1\atop k\not=i}^N \frac{1}{\xi_k - \xi_i}\Biggr)
    \Biggl(\prod_{\beta=1}^{N-1}
	\frac{\lambda_{\beta} - \xi_i}{\lambda_N - \lambda_{\beta}}\Biggr)
    \Bigg\lbrace
    \prod_{k=1\atop k\not=i}^N (\lambda_N - \xi_k + \eta)\\
    - \sum_{\beta=1}^{N-1} \prod_{k=1\atop k\not=i}^N
	(\lambda_{\beta} - \xi_k + \eta)\
	\frac{\lambda_N - \xi_i}{\lambda_{\beta} - \xi_i}\
	\Biggl(\prod_{\alpha=1\atop \alpha\not=\beta}^{N-1}
		\frac{\lambda_N - \lambda_{\alpha}}
		{\lambda_{\beta} - \lambda_{\alpha}}\Biggr) \Bigg\rbrace
   \ Z_{N-1}\left(\lbrace \lambda_{\alpha} \rbrace_{\alpha\not=N},
   \lbrace \xi_j \rbrace_{j\not=i} \right)\ ,
\end{multline*}
and one can easily see that
\begin{multline*}
  \prod_{j=1\atop j\not=i}^N (\xi_i - \xi_j + \eta) =
  \Biggl( \prod_{\beta=1}^{N-1} \frac{\xi_i - \lambda_{\beta}}
	{\lambda_N - \lambda_{\beta}} \Biggr)
  \Bigg\lbrace \prod_{k=1\atop k\not=i}^N (\lambda_N - \xi_k + \eta)\\
  - \sum_{\beta=1}^{N-1} \prod_{k=1\atop k\not=i}^N
	(\lambda_{\beta} - \xi_k + \eta)\
	\frac{\lambda_N - \xi_i}{\lambda_{\beta} - \xi_i}\
	\Biggl( \prod_{\alpha=1\atop \alpha\not=\beta}^{N-1}
	\frac{\lambda_N - \lambda_{\alpha}}{\lambda_{\beta} - \lambda_{\alpha}}
	\Biggr) \Bigg\rbrace\ ,
\end{multline*}
equality between two polynomials of degree $N-1$ in $\xi_i$,
which can be proved at the $N$ points $\xi_i = \lambda_{\alpha}$,
$1 \le \alpha \le N$.

Thus, as $\mathcal{N}$ et $\hat{\mathcal{N}}$ have the same determinant,
this finishes our proof of the formula~\eqref{partition} for the partition function.

\section*{Appendix B}
\renewcommand{\theequation}{B.\arabic{equation}}
\setcounter{equation}{0}

For the calculation of scalar products and correlation functions in the
$F$-basis, we need determinant
 representations for the following  functions 
\begin{align} 
  G_B^{(0)}(\{\la_k\}, i_1,\dots,i_n) &\equiv\bra{i_1,\dots,i_n}\
   \prod_{k=1}^n \Bt (\lambda_k)\ \ket{0},\label{partB}\\
   G_C^{(0)}(\{\mu_l\}, i_1,\dots,i_n) &\equiv\bra{0}\ 
    \prod_{l=1}^n \Ct (\mu_l)\ \ket{i_1,\dots,i_n}.
\label{partC}
\end{align}
Such representations can be easily obtained 
 from the one of the partition function $Z_n$,
by calculating explicitly the action of the operators
$\Bt_{1\ldots N}(\lambda_\alpha)$ or $\Ct_{1\ldots N}(\mu_\alpha)$ in the sites which do not
 belong
to $\lbrace i_1,\ldots,i_n \rbrace$.
Indeed, in the expression for the operator $B(\la)$ in the $F$-basis
\begin{equation}
  \Bt_{1\ldots N} (\lambda_\alpha)\ =\ \sum_{i = 1}^{N}\ \sigma_i^-\ 
    c(\lambda_\alpha,\xi_i)\ 
  \tens_{k \ne i}\ 
  \left( 
   \begin{array}{cc}
       b (\lambda_\alpha, \xi_k) & 0\\
       0 & b^{-1}(\xi_k,\xi_i)
   \end{array}
  \right)_{[k]},\label{BTbaseF}
\end{equation}
only the terms corresponding to a $\sigma_{i_k}^-$, where
$k$ belongs to $\lbrace 1,\ldots,n \rbrace$, give a
non-zero contribution to the function  $G_B^{(0)}(\{\la_k\}, i_1,\dots,i_n)$.
It means that the operators $\Bt_{1\ldots N} (\lambda_\alpha)$,
$1 \le \alpha \le n$, act  as diagonal ones on all the spaces
except $i_1,\ldots,i_n$. Thus, we can extract the action of 
$\prod_{\alpha=1}^{n} \Bt_{1\ldots N}(\lambda_\alpha)$ 
on all the  sites but
$i_1,\ldots,i_n$ as a global factor:
\begin{align}
 G_B^{(0)}(\{\la_k\}, i_1,\dots,i_n)
  &= \biggl(\prod_{\alpha=1}^n \prod_{k=1 \atop k \ne i_1,\ldots,i_n}^N 
       b(\lambda_{\alpha},\xi_k) \biggr)  
 Z_n \left(\lbrace \lambda_{\alpha} \rbrace, \lbrace \xi_{i_j}\rbrace \right)\nonumber\\
  &= \biggl(\prod_{\alpha=1}^n \prod_{k=1}^N 
         b(\lambda_{\alpha},\xi_k) 
     \biggr)
  \frac{\pl_{\alpha=1}^n\pl_{k=1}^n 
          \varphi\left(\lambda_{\alpha}-\xi_{i_k}+\eta\right)}
     {\pl_{j>k} \varphi\left(\xi_{i_j}-\xi_{i_k}\right)
     \pl_{\alpha<\beta} \varphi\left(\lambda_{\alpha}-\lambda_{\beta}\right)
     }\det \mathcal{N}(\lbrace \lambda_{\alpha} \rbrace, \lbrace \xi_{i_j}\rbrace)\ , \label{Bpart-partiel}
\end{align}
where $\mathcal{N}(\lbrace \lambda_{\alpha} \rbrace, \lbrace \xi_{i_j}\rbrace)$ is the $n \times n$ matrix 
defined by~\eqref{matrXXX} or~\eqref{matrXXZ}.

 Using the representation of the operator $C(\mu)$ in the $F$-basis one can analogously obtain a representation
 for the function $G_C^{(0)}(\{\mu_l\},i_1,\dots,i_n)$ \eqref{partC}. 
Both in the  XXX and XXZ cases it is given by
\begin{align}
G_C^{(0)}(\{\mu_l\},i_1,\dots,i_n)
= \biggl(\prod_{j=1}^n \prod_{k=1 \atop k \ne i_j}^N b^{-1}(\xi_{i_j},\xi_k)
     \biggr)
     \biggl( \prod_{j,k=1 \atop j\ne k}^n b(\xi_{i_j},\xi_{i_k}) \biggr)G_B^{(0)}(i_1,\dots,i_n,\{\mu_l\}). 
\label{Cpart-partiel}
\end{align}

\section*{Appendix C}
\renewcommand{\theequation}{C.\arabic{equation}}
\setcounter{equation}{0}

In this Appendix we calculate recursively the intermediate 
functions $\G|^{(m)}$~(\ref{interfun}). The detailed proof is written here  
only for the XXX case, but is quite
similar in the XXZ case.

 To compute the function  $\G|^{(1)}$ starting from the expression~\eqref{G0}
for $\G|^{(0)}$, one has to perform the summation in~\eqref{recur1}. 
Taking into account
that only the
first column of the matrix $\mathcal{N}$ depends on the parameter
$\xi_{i_1}$, one can express $\G|^{(1)}$ as follows
\begin{equation}
\G|^{(1)}(\{\la_\a\},\mu_1,i_{2},\dots,i_n)=
  \frac{\pl_{\alpha=1}^n\pl_{k=2}^n \left(\lambda_\a-\xi_{i_k}+\eta\right)}
     {\pl_{n\geq j>k\geq 2} \left(\xi_{i_k}-\xi_{i_j}\right)
     \pl_{1\leq\alpha<\beta\leq n} \left(\lambda_{\b}-\lambda_{\a}\right)
     }\det \mathcal{N}^{(1)}(\lbrace \lambda_{\a} \rbrace,\mu_1,i_{2},\dots,i_n),
\label{g1}
\end{equation}
where the matrix 
$\mathcal{N}^{(1)}(\lbrace \lambda_{\alpha} \rbrace,\mu_1,i_{2},\dots,i_n)$
is defined as
\begin{align}
  \mathcal{N}^{(1)}_{a b} &=\frac 1{\mu_1-\xi_{i_b}}\mathcal{N}_{a b}\qquad 
    \text{for}\quad b\geq 2,\\
  \mathcal{N}^{(1)}_{a 1} &=\pl_{k=2}^n(\mu_1-\xi_{i_k}+\eta)\sul_{i_1=1}^n
            \frac \eta{\mu_1-\xi_{i_1}}\frac \eta{\la_a-\xi_{i_1}}
            \frac{\pl_{m\neq a}(\la_m-\xi_{i_1}+\eta)}
                 {\pl_{k=2}^n(\xi_{i_1}-\xi_{i_k}+\eta)}
            \pl_{j\neq i_1}b^{-1}(\xi_{i_1},\xi_j).
\label{recur2}
\end{align}
It should be mentioned that the summation in \eqref{recur2} can be taken over all the possible
values of $i_1$ as the contributions of the terms $i_1=i_j,\quad j>1$ are equal to zero. It is
possible to calculate explicitly the sum in  \eqref{recur2} using its analytical properties and
the fact that $\{\la_\a\}$ is a solution of Bethe equations:
\begin{multline}
     \sul_{i_1=1}^n\frac \eta{\mu_1-\xi_{i_1}}\frac 
    \eta{\la_a-\xi_{i_1}}\frac{\pl_{m\neq a}(\la_m-\xi_{i_1}+\eta)}
                              {\pl_{k=2}^n(\xi_{i_1}-\xi_{i_k}+\eta)}
     \pl_{j\neq i_1}b^{-1}(\xi_{i_1},\xi_j)
         =\frac{H_{a1}}{\pl_{k=2}^n(\mu_1-\xi_{i_k}+\eta)}\\
          +\sul_{b=2}^n \frac{1}{\mu_1-\xi_{i_b}+\eta}
            \cdot\frac{\pl_{m=1}^n(\la_m-\xi_{i_b})}
        {\pl_{j=1\atop j\neq b}^n(\xi_{i_b}-\xi_{i_j})}\cdot
          \frac \eta{(\la_a-\xi_{i_b})(\la_a-\xi_{i_b}+\eta)},
\label{sumpol}
\end{multline}
 where the function $H_{a b}$ has the form
\begin{equation}
  H_{a b}=\frac{\eta}{\la_a-\mu_b}
      \Bigl(r(\mu_b)\pl_{m\neq a}(\la_m-\mu_b+\eta)
            -\pl_{m\neq a}(\la_m-\mu_b-\eta)\Bigr),
\label{matrslav}
\end{equation}
with 
\[r(\mu)=\frac{a(\mu)}{d(\mu)}=\pl_{j=1}^N\frac{\mu-\xi_j+\eta}{\mu-\xi_j}.\]
Indeed, the left hand side of \eqref{sumpol} is a 
rational function of $\mu_1$ with simple poles at the points 
$\mu_1=\xi_j, \ j=1,\dots, N$, 
 and its limit is zero when $\mu_1\rightarrow\infty$. The right hand side is 
also a rational function of $\mu_1$, which has only simple poles and becomes zero  when $\mu_1\rightarrow\infty$. 
The residues of the r.h.s.\ at the points $\mu_1=\xi_j$ are the same as in the l.h.s. Thus one should only
prove that the r.h.s.\ has no other poles, namely when $\mu_1=\la_a$ and $\mu_1=\xi_{i_k}-\eta,\quad k=2,\dots, n$.
One can easily see that the residues  of the r.h.s.\ at the points  $\mu_1=\xi_{i_k}-\eta$ are equal to zero.
As $\{\la_\a\}$ is a solution of Bethe equations, the
residue at the point $\mu_1=\la_a$ is also equal to zero. 
Therefore the l.h.s and r.h.s of  \eqref{sumpol}
are rational functions having the same behavior when $\mu_1\rightarrow\infty$, the same simple 
poles and the same residues in these poles, thus they are equal.

So, the matrix elements of the first column of the matrix 
$\mathcal{N}^{(1)}$ have the form
\[\mathcal{N}^{(1)}_{a 1}=H_{a 1}+\sul_{b=2}^n\alpha_b\mathcal{N}^{(1)}_{a b},\]
where $\a_b$ are  coefficients which do not depend on $a$.
Only the first term in this sum gives a nonzero contribution to the determinant of $\mathcal{N}^{(1)}$, which leads to 
the following representation for  $\G|^{(1)}$:
\begin{equation}
  \G|^{(1)}(\{\la_\a\},\mu_1,i_{2},\dots,i_n)=
   \frac{\pl_{\alpha=1}^n\pl_{k=2}^n \left(\lambda_\a-\xi_{i_k}+\eta\right)}
     {\pl_{n\geq j>k\geq 2} \left(\xi_{i_k}-\xi_{i_j}\right)
     \pl_{1\leq\alpha<\beta\leq n} \left(\lambda_{\beta}-\lambda_{\alpha}\right)
     }\det \mathcal{G}^{(1)}(\lbrace \lambda_{\alpha} \rbrace,\mu_1,i_{2},\dots,i_n),
\label{g1r}
\end{equation}
with the matrix $\mathcal{G}^{(1)}$  defined as
\begin{equation}
\begin{aligned}
  \mathcal{G}^{(1)}_{a b} &=\frac 1{\mu_1-\xi_{i_b}}\mathcal{N}_{a b}
            \qquad \text{for}\quad b\geq 2,\\
  \mathcal{G}^{(1)}_{a 1} &=H_{a 1}.
\end{aligned}
\end{equation}

Repeating this procedure, one obtains a general expression 
for all the functions  $\G|^{(m)}$:
\begin{multline}
  \G|^{(m)}(\{\la_k\},\mu_1,\dots,\mu_m,i_{m+1},\dots,i_n)=
     \frac{\pl_{\alpha=1}^n\pl_{k=m+1}^n \left(\lambda_\a-\xi_{i_k}+\eta\right)}
     {\pl_{n\geq j>k\geq m+1} \left(\xi_{i_k}-\xi_{i_j}\right)
     \pl_{1\leq\alpha<\beta\leq n} \left(\lambda_{\b}-\lambda_{\a}\right)
     }\times\\
  \times\frac 1{\pl_{m\geq j>k\geq 1} \left(\mu_k-\mu_j\right)}
    \det \mathcal{G}^{(m)}(\lbrace \lambda_{\alpha} 
   \rbrace,\mu_1,\dots,\mu_m,i_{m+1},\dots,i_n),
\label{gm}
\end{multline}
with the matrix  
$\mathcal{G}^{(m)}(\lbrace \lambda_{\alpha} \rbrace,\mu_1,\dots,\mu_m,i_{m+1},\dots,i_n)$
given by
\begin{alignat}{2}
   \mathcal{G}^{(m)}_{a b} &=\mathcal{N}_{a b}
               \pl_{j=1}^m\left(\frac 1{\mu_j-\xi_{i_b}}\right) & 
           \qquad &\text{for}\quad b>m,\\
   \mathcal{G}^{(m)}_{a b} &=H_{a b} & 
           \qquad &\text{for}\quad b\leq m. \label{matg}
\end{alignat}

 We prove~\eqref{gm}-\eqref{matg} by induction. For $m=1$ it coincides 
with~\eqref{g1r}. 
Let  this representation be valid for $\G|^{(m-1)}$. Combining it
with \eqref{recur1} and \eqref{Celem} we obtain 
the following expression for $\G|^{(m)}$: 
\begin{multline}
   \G|^{(m)}(\{\la_k\},\mu_1,\dots,\mu_m,i_{m+1},\dots,i_n)=
     \frac{\pl_{\alpha=1}^n\pl_{k=m+1}^n \left(\lambda_\a-\xi_{i_k}+\eta\right)}
     {\pl_{n\geq j>k\geq m+1} \left(\xi_{i_k}-\xi_{i_j}\right)
     \pl_{1\leq\alpha<\beta\leq n} \left(\lambda_{\b}-\lambda_{\a}\right)
     }\times\\
   \times\frac 1{\pl_{m-1\geq j>k\geq 1} \left(\mu_k-\mu_j\right)}
       \det \mathcal{N}^{(m)}(\lbrace \lambda_{\alpha} 
          \rbrace,\mu_1,\dots,\mu_m,i_{m+1},\dots,i_n),
\label{recurm}
\end{multline}
where the matrix $\mathcal{N}^{(m)}(\lbrace \lambda_{\alpha} \rbrace,\mu_1,\dots,\mu_m,i_{m+1},\dots,i_n)$
is defined as
\begin{alignat}{2}
   \mathcal{N}^{(m)}_{a b} &=\mathcal{N}_{a b} 
                \pl_{j=1}^m\left(\frac 1{\mu_j-\xi_{i_b}}\right) &
            \qquad &\text{for}\quad b>m,\nonumber\\
   \mathcal{N}^{(m)}_{a b} &=H_{a b} &
            \qquad &\text{for}\quad b< m,\nonumber\\
   \mathcal{N}^{(m)}_{a m} &=\pl_{k=m+1}^n(\mu_1-\xi_{i_k}+\eta)
                  \sul_{i_m=1}^n \frac \eta{\mu_m-\xi_{i_m}}
                  \frac \eta{\la_a-\xi_{i_m}}\times\nonumber\\
       &\qquad\times\frac{\pl_{l\neq a}(\la_l-\xi_{i_m}+\eta)}
                         {\pl_{k=m+1}^n(\xi_{i_m}-\xi_{i_k}+\eta)
                          \pl_{j=1}^{m-1}(\mu_j-\xi_{i_m})}
            \pl_{j\neq i_m}b^{-1}(\xi_{i_m},\xi_j).
\label{recurNm}
\end{alignat}
 The sum in~\eqref{recurNm} can be computed the same way 
as in~\eqref{sumpol}. One can prove
using similar arguments that
\begin{equation}
   \mathcal{N}^{(m)}_{a m}=\frac{H_{a m}}
        {\pl_{j=1}^{m-1}(\mu_j-\mu_m) \pl_{k=m+1}^n (\mu_m-\xi_{i_k}+\eta)}
        +\sul_{b=1}^{m-1}\b^{(m)}_b H_{a b} 
        +\sul_{b=m+1}^n\a^{(m)}_b\mathcal{N}_{a b},
\end{equation}
where $\alpha^{(m)}_b$ and $\b^{(m)}_b$ are coefficients which do not depend on $a$.
As only the first term gives a nonzero contribution to the determinant we obtain the representation
\eqref{gm}.

Finally the scalar product is given by 
\begin{equation}
   \S|_n(\{\mu_j\},\{\la_\a\})=\G|^{(n)}(\{\la_\a\},\mu_1,\dots,\mu_n)
            =\frac {\det H(\{\la_\a\},\{\mu_j\})}
              {\pl_{j>k} \left(\mu_k-\mu_j\right)
     \pl_{\alpha<\beta} \left(\lambda_{\b}-\lambda_{\a}\right)},
\label{slavXXX1}
\end{equation}
with matrix elements of $H(\{\la_\a\},\{\mu_j\})$ defined by \eqref{matrslav}.

\bibliographystyle{h-elsevier} 

\bibliography{biblio}


\end{document}